\documentclass[]{aa}
\usepackage{graphicx}
\usepackage{txfonts}
\begin{document}
\title{Non-adiabatic theoretical observables in $\delta$ Scuti stars}

\author{A. Moya \inst{1} \and R. Garrido \inst{1} \and M.A. Dupret \inst{1,2}
\thanks{Marie Curie Postdoctoral Fellow, European Union}}

\offprints{A. Moya
\email{moya@iaa.es}   }

\institute{ Instituto de Astrof\'{\i}sica de Andaluc\'{\i}a-CSIC, Granada,
Spain\\
\and
Institut d'Astrophysique et de G\'eophysique de l'Universit\'e de Li\`ege,
Belgium}

\abstract{Phase differences and amplitude ratios at different colour 
photometric bands are currently being used to discriminate pulsation modes
 in order to facilitate mode identification
of $\kappa$-driven non-radial pulsating stars. In addition to physical inputs 
(e.g., mass, $T_{eff}$, etc.), these quantities depend on the non-adiabatic 
treatment of the atmosphere. This paper presents theoretical results concerning
$\delta$ Scuti pulsating stars. The envelope of each of these stellar 
structures possesses a convection zone whose development is determined
by various factors. An interacting pulsation-atmosphere physical treatment 
is introduced which supplies two basic non-adiabatic physical quantities: 
 the relative effective temperature variation and
the phase lag $\phi^T$, defined as the angle between effective temperature 
variations and radial displacement. These quantities can be used to derive 
the phase differences and amplitude ratios. Numerical
values for these quantities depend critically on the $\alpha$ MLT parameter 
used to calculate the convection in the envelope. The dependence on $\alpha$ 
was analized and it was found that the use of colour observations may be of 
considerable importance in testing the MLT.
Finally, examples are given of how $\alpha$ introduces uncertainties 
in the theoretical predictions regarding phases and amplitudes of photometric 
variations in $\delta$ Scuti pulsating stars.

}

\maketitle


\section{Introduction}

Asteroseismology is presently being developed as an efficient instrument in the
study of stellar interiors and evolution. Pulsational periods are the most 
important asteroseismological observational inputs. However,
for non-solar like oscillations, such as those characterizing $\delta$ 
Scuti stars, the information contained in these periods is not sufficient for 
adequate constraints in theoretical predictions. Without additional 
observational data, mode identification is, therefore, not feasible in 
non-solar like contexts. The primary obstacles 
are that the pulsation modes of $\delta$ Scuti stars: 1) do not lie within the 
asymptotic regime and 2) may be affected by the ``avoided crossing'' 
phenomenon. Furthermore, rotation and eventual coupling also destroy
any possible regular pattern.

One way to obtain more information on the basis of photometric observations is 
to study the multicolor flux variations. The linear approximation to non-radial
 flux variations of a pulsating star was 
first derived by Dziembowski (\cite{Dziem77b}), and later reformulated by 
Balona and Stobie (\cite{Bal79a}, \cite{Bal79b}) and by 
Watson (\cite{Wat88}). In Garrido et al. (\cite{Rafa90}), the linear
approximation was successfully used to discriminate the spherical 
orders $l$ of these pulsating star modes. Several attempts to
fit observations have shown that the method can be applied, 
at least to low rotational velocities (e.g., FG Vir in Breger et al. 
\cite{Bre99}; BI CMi in Breger et al. \cite{Bre02a}; 4 CVn in Breger et al. 
\cite{Bre02b}; V1162 in Arentoft et al. \cite{Are02} and some other stars
 in Garrido \cite{Rafa00}; for fast rotators, see Daszy\'nska et al. 
\cite{Dasz02}).

Most theoretical and numerical pulsation models have been developed using the
adiabatic approximation (Christensen-Dalsgaard \cite{JCD}; Tran \& Leon 
\cite{FILOU}). However, pulsation is highly non-adiabatic in 
stellar surface layers, in which thermal relaxation time is either of the same 
order or even lower than the pulsation period. The accurate determination 
of the eigenfunctions in these layers will therefore require the use of a 
non-adiabatic description which includes the entire atmosphere. This procedure
then makes it possible to relate multicolor photometric observables with such 
eigenfunctions.

A number of authors (e.g., Dziembowski \cite{Dziem77a}; Saio and Cox 
\cite{Saio80}; Pesnell \cite{Pes90}; Townsend \cite{Townsend02}) have developed
non-adiabatic codes and performed stellar pulsation calculations. However,
these works have been carried out without a complete description of 
pulsation-atmosphere interaction.

In this paper two approaches are compared. The first follows Unno et al. 
(\cite{Unno}) for the equations and the numerical method used to solve them. 
Outer boundary conditions are imposed at the level of the photosphere. This 
approximation will be referred to below as ``without atmosphere''. The second
applies a non-adiabatic pulsational treatment to the atmosphere derived by 
Dupret et al. (\cite{MA02}). This treatment has made it possible to
 obtain photometric observable values that are more realistic when compared 
with those generated by the ``without'' atmosphere approximation (referred to 
below as ``with atmosphere''). The latter approach allows theoretical 
predictions to be more directly connected with photometric observations than 
they previously had been by using only period comparisons.

Theoretical and observed amplitude ratios and phase differences
 can be compared as observed in different color photometric bands.
Garrido et al. (\cite{Rafa90}) showed that the wavelength dependence of the 
limb darkening integrals is very weak for low $l$ values. Thus, 
at least three color combinations, distributed in the widest possible
 range of wavelengths, give consistent values for phase lag 
$\phi^T$ and $R$. The latter parameter was defined by Watson (\cite{Wat88}) and
designed to measure departures from adiabatic conditions. Comparisons 
between non-adiabatic predictions and multicolor photometric observations have 
been made by Cugier et al. (\cite{cug}) for $\beta$ Cephei stars, 
Balona \& Evers (\cite{balev}) for $\delta$ Scuti stars and Townsend 
(\cite{Townsend02}) for Slowly Pulsating B stars. However, such comparisons
were made without carrying out a detailed non-adiabatic pulsation 
treatment for the atmosphere. Photometric magnitude variations have been 
reformulated by Dupret et al. (\cite{MA03}), thus making it possible to obtain 
theoretical non-adiabatic quantities which explicitly include the atmosphere 
(Dupret et al. \cite{MA02}). This reformulation has been 
employed in the study of $\beta$ Cephei and Slowly Pulsating B stars.

In the present paper the approach of Dupret et al. (\cite{MA02}, \cite{MA03}) 
is applied to $\delta$ Scuti stars. Non-adiabatic theoretical
models ``with'' and ``without'' atmosphere, as well as their resolution 
algorithms, are presented. Solutions are compared in order to show the 
improvement introduced by including the pulsation-atmosphere interaction. One 
possible procedure for mode identifications using multicolor 
photometric diagrams is also suggested. For $\delta$ Scuti
 stars, the non-adiabatic observables with which we are working can be 
directly related with the characteristics of their thin convective envelopes. 
These envelopes are described by the Mixing Length Theory (MLT) and 
parametrized by $\alpha$, which is defined as the proportionality constant 
relating the mean path of a convective element with local pressure scale 
height. The $\alpha$ parameter can be constrained by searching for the best fit
between theoretical and observed photometric amplitude ratios and phase 
differences in color photometric bands. Such observables also depend on the 
opacity variation in the HeII, HeI and HI ionization zones.


\section{Theoretical models}

Equilibrium stellar models have been computed using the CESAM
 code (Morel \cite{Morel97}). In accordance with this code, the stellar 
atmosphere can be described: 1) as a single layer (i.e., the 
photosphere) calculated in the Eddington approximation, or 2) using the Kurucz 
equilibrium atmospheric models (Kurucz \cite{Kurucz}) to reconstruct the 
atmosphere from a specific Rosseland optical depth until reaching the last
edge of the 
star. Table \ref{carac} shows the global characteristics of the equilibrium
 models. In the convective core of these models, overshooting 
has been set to 0.2 times the local pressure scale height. 

\begin{table}
\begin{tabular}{rrrrrr} \hline
   \noalign{\medskip}
${M\over M_{\sun}}$ & $\log T_{\rm eff}$ & $\log g$ & $\log {L\over L_{\sun}}$ & 
$\log {R\over R_{\sun}}$ & $X_c$ \\ 
   \noalign{\smallskip}
\hline 
   \noalign{\smallskip}
2.0 & 3.941 & 4.181 & 1.273 & 1.900 & 0.55 \\
2.0 & 3.879 & 3.845 & 1.367 & 2.799 & 0.25 \\
1.8 & 3.889 & 4.064 & 1.138 & 2.063 & 0.44 \\ 
   \noalign{\smallskip}
\hline
   \noalign{\smallskip}
\end{tabular}
\caption{Global characteristics of the equilibrium models}
\label{carac}
\end{table}

The pulsational code begins by computing the adiabatic solution for fixed 
values of $n$ and $l$, and this computation is used as trial input 
for the non-adiabatic computations ``with'' or ``without'' atmosphere. 
These calculations then make it possible to derive $\phi^T$,  
$|\delta\mathrm{T}_{\rm eff}/\mathrm{T}_{\rm eff}|$ and $\delta g_e/g_e$, which
are directly related with the photometric color variations.

\subsection{Non-adiabatic models ``without'' atmosphere}

Here the adiabatic and non-adiabatic equations have been derived following 
Unno et al. (\cite{Unno}). However, two variables regarding the stellar 
interior have been modified in order to directly link the interior and 
atmosphere pulsation equations. These variables are

\begin{eqnarray}
	y_1 = {\xi_{\, r} \over r} ~~~~~~~~~~\;\;\;
	y_2 = {\delta P_g \over P_g} ~~~~~~~~~~\;\;\;
	y_3 = {\Phi^{\prime} \over gr} ~~~~~~~~~~\;\;\;
\nonumber
\\
	y_4 = {d \, \Phi^{\prime} \over g \, dr } ~~~~~~~~~~\;\;\;
	y_5 = {\delta T \over T} ~~~~~~~~~~\;\;\;
	y_6 = {\delta L_R \over L_R} ~~~~~~~~~~\;\;\;
\end{eqnarray}

\noindent where $\xi_{\, r}$ is the radial displacement, $P_g$,
the gas pressure, $\Phi$, the gravitational potential, 
$L_R$, the radiative luminosity and $\delta X$ (resp. $X^{\prime}$), the 
Lagrangian (resp. Eulerian) perturbation of the $X$ variable.
The other symbols bear their usual meaning.

The linear non-adiabatic pulsation equations corresponding to these variables 
are described in Appendix A (Eqs. (\ref{eq1}) to (\ref{eq6})), and
 are obtained by neglecting the effects of rotation and the magnetic field.
 The "frozen" convection flux approximation, $\delta L_C = 0$ and 
$\vec{F}^{\prime}_{C_\perp} = 0$, has been chosen by following Unno et al. 
(\cite{Unno}). Boundary conditions are also taken from the same source.
Internal conditions are described by Eqs. (\ref{intmbc}) to (A.11) in 
Appendix A. Surface conditions are given here because they are particularly
 relevant to the comparison of the ``with atmosphere''
(Sect. \ref{withatmo}) and ``without atmosphere'' treatment results:

\begin{enumerate}
	\item  The mechanical boundary condition
\vspace{0.1cm}
	\begin{eqnarray}
		& & \Big({l(l+1)\over\omega^2}-4-\omega^2\Big)\,y_1-
		\beta\,y_2+
\nonumber
\\
		& & \Big({l(l+1)\over\omega^2}-1-l\Big)\,y_3-
		{4\over 3}{a\mathrm{T}^4\over\mathrm{P}}\,y_5 = 0
	\label{extmbc}
	\end{eqnarray}
\vspace{0.1cm}
	\item The potential boundary condition
\vspace{0.1cm}
	\begin{equation}
		(l+1)\,y_3+y_4 = 0
	\label{extpbc}
	\end{equation}
\vspace{0.1cm}
	\item The thermal boundary condition
\vspace{0.1cm}
	\begin{eqnarray}
		2\,y_1+4\,y_5-y_6 = 0
	\label{exttbc}
	\end{eqnarray}
\end{enumerate}
%

\noindent Note that Eq. (\ref{exttbc}) is obtained by perturbing the equation 
which defines the effective temperature 
($L_R=4\pi R^2 \sigma_{rad}\,T_{\rm eff}^4$), as well as by assuming that in 
the photosphere:

\begin{equation}
\frac{\delta T}{T}= \frac{\delta T_{\rm eff}}{T_{\rm eff}}
\end{equation}

\subsection{Non-adiabatic models ``with'' atmosphere}
\label{withatmo}

Eqs. (\ref{eq1})-(\ref{eq6}) are also used to describe the pulsation inside 
the star, whereas Eqs. (\ref{intmbc})-(A.11) describe the internal boundary 
conditions. However, the complete atmosphere pulsation equations
 are solved following Dupret et al. (\cite{MA02}). The latter equations 
have been derived by taking into account that: 1) the radiation field is not 
isotropic in the atmosphere and the diffusion approximation is therefore not 
valid in this case; and 2) the radiation stress tensor cannot be represented by
 a diagonal matrix with a constant element: 
$\mathrm{P}_R = (1 / 3) a\mathrm{T}^4$.

For these reasons, new approximations have been used with respect to those 
solved for the interior in order to obtain an improved description of the 
atmosphere. This new approach is based on the fact that the atmosperic thermal 
relaxation time is very short as compared to pulsation periods. Consequently, 
it can be assumed that during each pulsation period, the atmosphere remains in 
thermal equilibrium. Rewriting the equations given in Dupret et al. 
(\cite{MA02}), those corresponding to the atmosphere are:

\begin{eqnarray}
	x{dy_1 \over dx} & = &
		\Big({l(l+1)\over c_1\omega^2}-3\Big)\,y_1+
		\Big({l(l+1)\over c_1\omega^2}{P_g\over \rho gr}-{1\over
		\mathrm{P}_{g \rho}}\Big)\,y_2 +
\nonumber
\\
	&&
		{l(l+1)\over c_1\omega^2}\,y_3 +
		\Big({\mathrm{P}_{g \mathrm{T}}\over \mathrm{P}_{g \rho}}+
		{l(l+1)\over c_1\omega^2}{\mathrm{a}_R\over \nabla g V}
		\Big)\,y_5
\label{eqma1}
\\
	x{dy_2 \over dx} & = &
		{\rho gr\over P_g}\Bigg(c_1\omega^2-\mathrm{U}+2+
		\Big({\mathrm{a}_R\over g}-1\Big)\Big({l(l+1)\over c_1\omega^2}
		-2\Big)\Bigg)\,y_1 +
\nonumber
\\
	&&
		{\rho gr\over P_g}\Bigg(\Big({\mathrm{a}_R\over g}-1\Big)
		\Big({l(l+1)\over c_1\omega^2}{P_g\over \rho gr}-1\Big)+
		{\mathrm{a}_R\over g}{\kappa_{\rho}\over \mathrm{P}_{g \rho}}
		\Bigg)\,y_2 +
\nonumber
\\
	&&
		\Big({\mathrm{a}_R\over g}-1\Big){l(l+1)\over c_1\omega^2}
		{\rho gr\over P_g}\,y_3 -
		{\rho gr\over P_g}\,y_4 +
\nonumber
\\
	&&
		{\rho gr\over P_g}\Bigg(\Big({\mathrm{a}_R\over g}-1\Big)
		{l(l+1)\over c_1\omega^2}{\mathrm{a}_R\over \nabla g V}+
		{\mathrm{a}_R\over g}\Big(\kappa_{\mathrm{T}}-
		\kappa_\rho{\mathrm{P}_{g \mathrm{T}}\over \mathrm{P}_{g \rho}}
		\Big)\Bigg)\,y_5 +
\nonumber
\\
	&&
		4{\rho r\mathrm{a}_R\over P_g}{\delta \mathrm{T}_{\rm eff}
		\over \mathrm{T}_{\rm eff}}
\label{eqma2}
\\
	x{dy_3 \over dx} & = &
		\big(1-\mathrm{U}\big)\,y_3+y_4
\label{eqma3}
\\
	x{dy_4 \over dx} & = &
		-\mathrm{U}{d\ln\rho\over d\ln r}\,y_1 +
		{\mathrm{U}\over\mathrm{P}_{g \rho}}\,y_2 +
		l(l+1)\,y_3 -
\nonumber
\\
	&&
		\mathrm{U}\,y_4 -
		\mathrm{U}{\mathrm{P}_{g \mathrm{T}}\over \mathrm{P}_{g \rho}}
		\,y_5
\label{eqma4}
\\
	x{dy_5 \over dx} & = &
		-{\kappa\rho r \over \tau}{\partial\ln\mathrm{T}\over
		\partial\ln\tau}\Big({l(l+1)\over c_1\omega^2}-2\Big)\,y_1 -
\nonumber
\\
	&&
		{\kappa\rho r \over \tau}{\partial\ln\mathrm{T}\over
		\partial\ln\tau}\Big({\kappa_{\rho}\over \mathrm{P}_{g \rho}}
		+{l(l+1)\over c_1\omega^2}{P_g\over \rho gr}\Big)\,y_2 -
\nonumber
\\
	&&
		{\kappa\rho r \over \tau}{\partial\ln\mathrm{T}\over
		\partial\ln\tau}{l(l+1)\over c_1\omega^2}\,y_3 -
\nonumber
\\&&
		{\kappa\rho r \over \tau}\Bigg({\partial\ln\mathrm{T}\over
		\partial\ln\tau}\Big(\kappa_{\mathrm{T}}-\kappa_{\rho}
		{\mathrm{P}_{g \mathrm{T}}\over \mathrm{P}_{g \rho}}+
		{l(l+1)\over c_1\omega^2}{\mathrm{a}_R\over \nabla g
		\mathrm{V}}\Big)-
\nonumber
\\
	&&
		1+\Big({\partial^2\ln\mathrm{T}\over
		\partial\ln\tau^2}\Big)\biggm/ \Big({\partial\ln\mathrm{T}\over
		\partial\ln\tau}\Big)\Bigg)\,y_5 -
\nonumber
\\&&
		{\kappa\rho r \over \tau}\Bigg[\bigg(1-\Big({\partial^2\ln
		\mathrm{T}\over\partial\ln\tau^2}\Big)\biggm/ \Big({\partial
		\ln\mathrm{T}\over\partial\ln\tau}\Big)\bigg)\bigg({\partial
		\ln\mathrm{T}\over\partial\ln\mathrm{T}_{\rm eff}}{\delta 
		\mathrm{T}_{\rm eff}\over\mathrm{T}_{\rm eff}}+
\nonumber
\\
	&&
		{\partial\ln\mathrm{T}
		\over\partial\ln\mathrm{g}_{e}}{\delta \mathrm{g}_{e}\over
		\mathrm{g}_{e}}\bigg)+{\partial^2\ln\mathrm{T}\over \partial
		\ln\tau\partial\ln\mathrm{T}_{\rm eff}}{\delta\mathrm{T}_{\rm eff}
		\over \mathrm{T}_{\rm eff}}+
\nonumber
\\
	&&
		{\partial^2\ln\mathrm{T}\over \partial
		\ln\tau\partial\ln\mathrm{g}_{e}}{\delta\mathrm{g}_{e}
		\over \mathrm{g}_{e}}\Bigg]
\label{eqma5}
\\
	x{dy_6 \over dx} & = &
		0
\label{eqma6}
\end{eqnarray}

\noindent where the new variables are defined as:

\begin{equation}
\mathrm{P}_g(\mathrm{or}\,\kappa)_{\rho}={\partial\ln P_g \,(\mathrm{or}\;
\kappa)\over\partial\ln\rho}\biggm|
_{\mathrm{T}}\;\;\;\;\mathrm{P}_g\,(\mathrm{or}\,\kappa)_{\mathrm{T}}={\partial
\ln P_g(\mathrm{or}\;\kappa)\over\partial\ln
\mathrm{T}}\biggm|_{\rho}
\end{equation}

\noindent and where $\mathrm{a}_R$ is the radiative acceleration and 
$\tau$ the Rosseland optical depth. The atmospheric equilibrium quantities 
for these equations have been obtained from the Kurucz model atmospheres 
(Kurucz \cite{Kurucz}).

In comparison to those used for the ``without'' atmosphere treatment, the 
boundary conditions for these equations are:

\begin{enumerate}
	\item The mechanical boundary condition is obtained by neglecting the 
derivative of the gas pressure perturbation at the stellar surface:
	\begin{equation}
		x {d\,y_2\over dx} = 0\,.
\label{lala}
	\end{equation}
	Inserting Eq. (\ref{lala}) in Eq. (\ref{eqma2}):
	\begin{eqnarray}
		\Bigg(c_1\omega^2-\mathrm{U}+2+
		\Big({\mathrm{a}_R\over g}-1\Big)\Big({l(l+1)\over c_1\omega^2}
		-2\Big)\Bigg)\,y_1 +
	&&
\nonumber
\\
		\Bigg(\Big({\mathrm{a}_R\over g}-1\Big)
		\Big({l(l+1)\over c_1\omega^2}{P_g\over \rho gr}-1\Big)+
		{\mathrm{a}_R\over g}{\kappa_{\rho}\over \mathrm{P}_{g \rho}}
		\Bigg)\,y_2 +
	&&
\nonumber
\\
		\Big({\mathrm{a}_R\over g}-1\Big){l(l+1)\over c_1\omega^2}\,y_3
		- y_4 +
&&
\nonumber
\\
		\Bigg(\Big({\mathrm{a}_R\over g}-1\Big)
		{l(l+1)\over c_1\omega^2}{\mathrm{a}_R\over \nabla g V}+
		{\mathrm{a}_R\over g}\Big(\kappa_{\mathrm{T}}-
		\kappa_\rho{\mathrm{P}_{g \mathrm{T}}\over \mathrm{P}_{g \rho}}
		\Big)\Bigg)\,y_5 +
	&&
\nonumber
\\
		4{\mathrm{a}_R\over g}{\delta \mathrm{T}_{\rm eff}
		\over \mathrm{T}_{\rm eff}}
		& = & 0
	\label{extmbcma}
	\end{eqnarray}
	\item The potential boundary condition remains the same:
	\begin{equation}
		(l+1)\,y_3+y_4 = 0
	\label{extpbcma}
	\end{equation}
	\item The thermal boundary condition, following Dupret et al. 
(\cite{MA02}):
	\begin{eqnarray}
		{\partial\ln\mathrm{T}\over \partial\ln\mathrm{T}_{\rm eff}}
		{\delta\mathrm{T}_{\rm eff}\over \mathrm{T}_{\rm eff}}+
		{\partial\ln\mathrm{T}\over \partial\ln\mathrm{g}_{e}}
		{\delta\mathrm{g}_{e}\over \mathrm{g}_{e}}+
	&&
\nonumber
\\
		{\partial\ln\mathrm{T}\over\partial\ln\tau}\Big({l(l+1)\over 
		c_1\omega^2}-2\Big)\,y_1+
	&&
\nonumber
\\
		{\partial\ln\mathrm{T}\over\partial\ln\tau}\Big({\kappa_{\rho}
		\over \mathrm{P}_{g \rho}}+{l(l+1)\over c_1\omega^2}{P_g \over
		\rho g r}\Big)\,y_2 +
	&&
\nonumber
\\
		{\partial\ln\mathrm{T}\over\partial\ln\tau}{l(l+1)\over 
		c_1\omega^2}\,y_3 +
\nonumber
\\
		\Bigg({\partial\ln\mathrm{T}\over\partial\ln\tau}\Big(
		\kappa_{\mathrm{T}}-\kappa_{\rho}{\mathrm{P}_{g\mathrm{T}}
		\over \mathrm{P}_{g\rho}}+{l(l+1)\over c_1\omega^2}
		{\mathrm{a}_R\over \nabla g \mathrm{V}}\Big)-1\Bigg)\,y_5
		& = & 0
	\end{eqnarray}
\end{enumerate}

A connecting layer is required which marks the border between the 
interior (Eqs. (\ref{eq1}) to (\ref{eq6})) and atmosphere treatments 
(Eqs. (\ref{eqma1}) to (\ref{eqma6})). The boundary
 conditions for the equilibrium models are calculated at the same connecting 
layer by imposing an adequate match with the Kurucz model atmosphere 
(Kurucz \cite{Kurucz}). Two restrictions are imposed in order to choose the 
location of this layer: 1) the diffusion approximation must be 
valid and 2) the layer must be located outside the convective envelope, since 
the atmospheric non-adiabatic treatment is not justified inside a convection 
zone, and because consistency requirements exist between the interior and 
atmosphere treatments.

\begin{figure}
\centering
	\resizebox{\hsize}{!}{\includegraphics{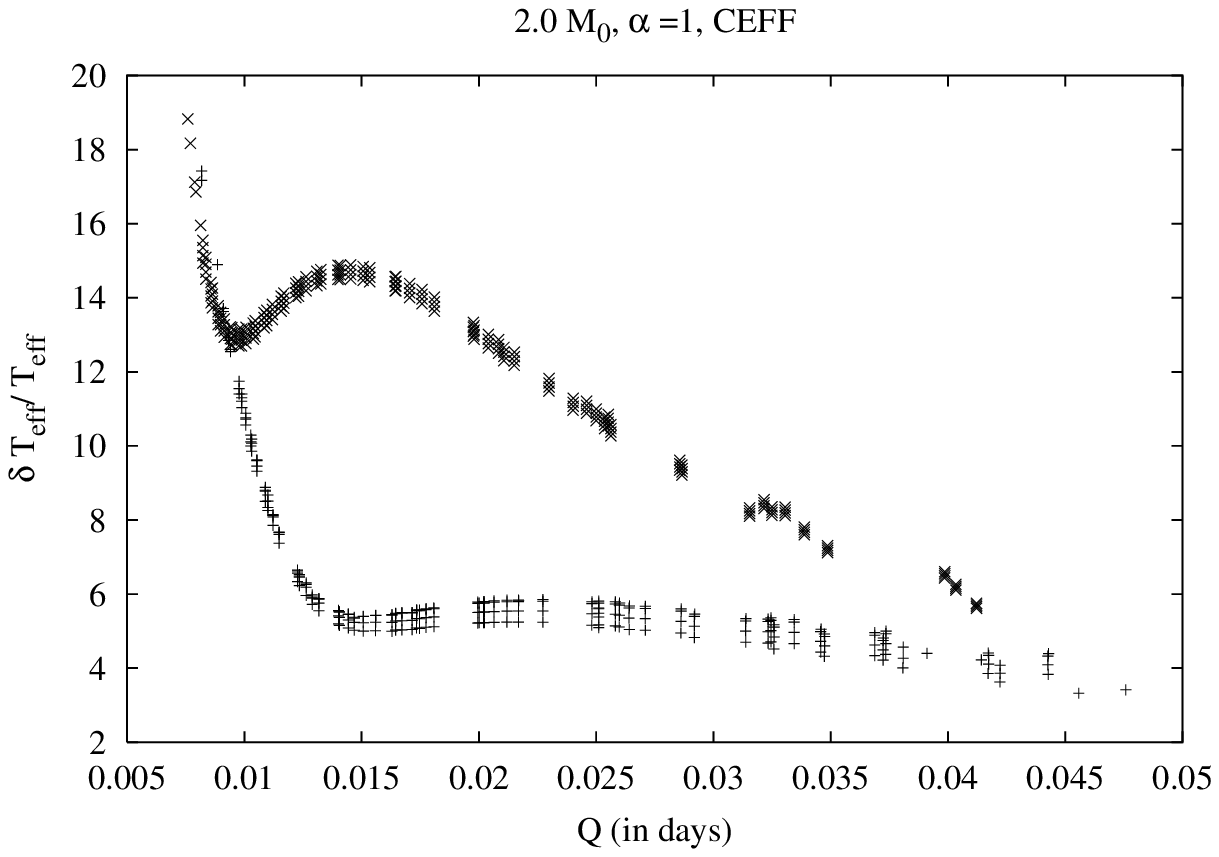}}
	\resizebox{\hsize}{!}{\includegraphics{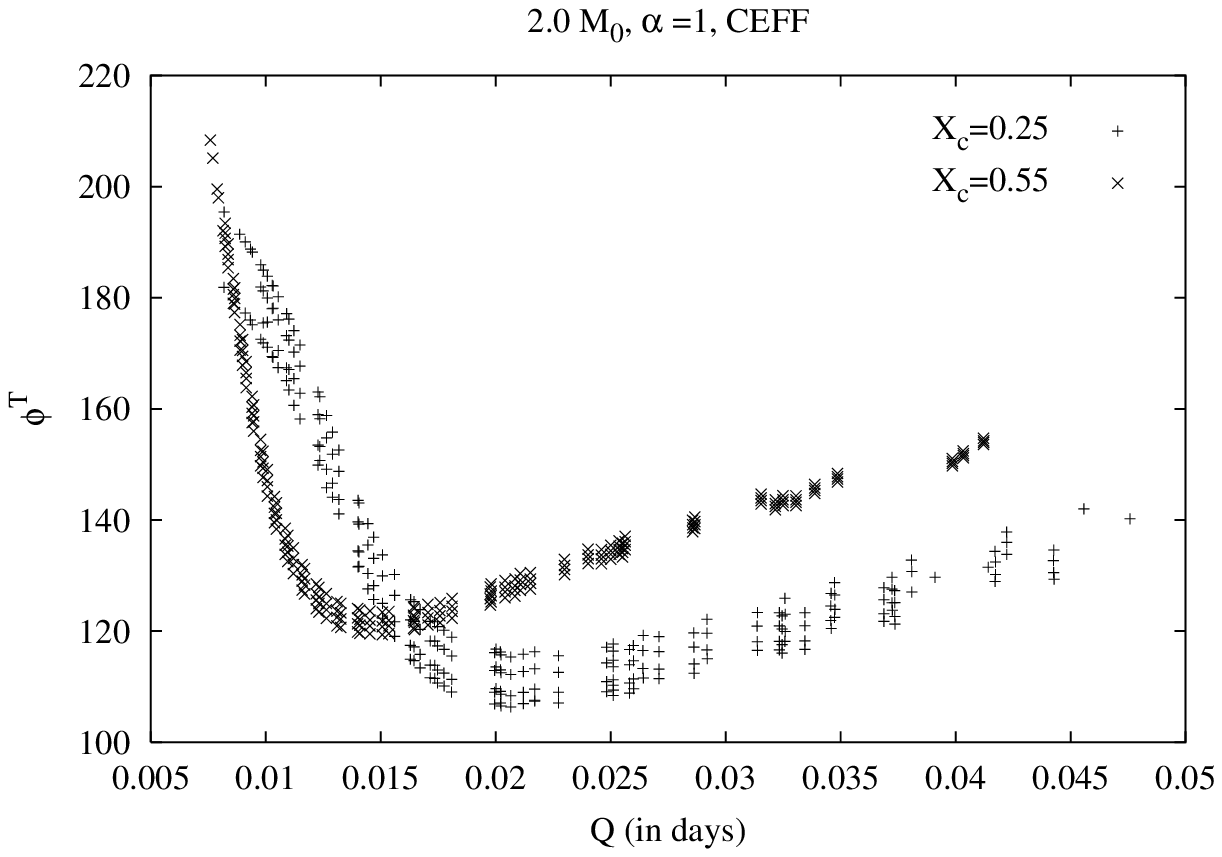}}
	\caption{Non-adiabatic quantities 
$|\delta\mathrm{T}_{\rm eff} / \mathrm{T}_{\rm eff}|$ (top) and $\phi^T$ 
(bottom), as a function of the pulsation constant $Q$ (in days) for different 
modes with spherical degrees $l=0,1,2,3$. Two $2.0 M_{\sun}$ models are solved 
at different evolutionary phases, with a MLT parameter $\alpha =1$ and the 
CEFF equation of state. Different values correspond to different locations of 
the connecting layer: $\tau=1,1.1,1.2,1.3$. ``$+$'' are for the model with 
$X_c=0.25$  and ``$\times$'' for $X_c=0.55$.
}
\label{est_tau}
\end{figure}

One of the uncertainties of the ``with'' atmosphere global algorithm is the 
location of the optical depth in which this connecting layer is defined, 
being that, despite the above-mentioned constraints, a small set of possible 
locations still exists. Fig. \ref{est_tau} shows the non-adiabatic 
results for two $2\,M_{\sun}$ stellar models at different evolutionary stages, 
locating the connecting layer at $\tau=1,1.1,1.2$ and $1.3$. Note that these 
results are not significantly affected by this choice. The connecting layer is 
selected at $\tau = 1$.

The two additional variables $\delta T_{eff}$ and $\delta g_e$ can be derived 
from:

\begin{eqnarray}
	4{\delta\mathrm{T}_{\rm eff}\over \mathrm{T}_{\rm eff}} & = &
		y_6-2\,y_1
\label{deltatt}
\\
	{\delta g_e \over g_e} & = &
		y_4+\Big(\mathrm{U}-2-c_1\omega^2\Big)\,y_1
\label{deltagg}
\end{eqnarray}

These quantities are determined at the photosphere 
($\log\mathrm{T}=\log\mathrm{T}_{\rm eff}$). The results 
are not significantly sensitive to the choice of this layer: a slight change 
of location implies changes which are always smaller than 5 percent for 
$|\delta\mathrm{T}_{\rm eff} / \mathrm{T}_{\rm eff}|$, smaller than 1 degree 
for the phase-lag $\phi^T$ and which are negligible for $\delta g_e / g_e$.

\section{Multicolor photometry}

An important application of our non-adiabatic code is that its theoretical 
predictions can be tested against multicolor photometric observations. The 
non-adiabatic quantities 
$|\delta\mathrm{T}_{\rm eff} / \mathrm{T}_{\rm eff}|$, $\delta g_e / g_e$ and
$\phi^T$ can be related to the photometric observables: amplitude ratios and 
phase differences between filters. In a one-layer linear approximation, the 
monochromatic magnitude variation of a non-radially pulsating star is given by:

\begin{eqnarray}
\label{deltaml}
	\delta m_{\lambda} & = &
%
		-{2.5\over \ln 10}\,a\, P_l^m\,(\cos i)\, b_{l\lambda}\,
		\big[-(l-1)\,(l+2)
		\cos\,(\sigma t) + 
\nonumber
\\
	&&
		f_{\mathrm{T}}\cos\,(\sigma t + 
		\phi^{\mathrm{T}})\,(\alpha_{\mathrm{T}\lambda} + 
		\beta_{\mathrm{T}\lambda}) 
\nonumber
\\&&
		- f_g \cos\,(\sigma t)\,(\alpha_{g \lambda} + 
		\beta_{g \lambda})\Big]
\end{eqnarray}
\noindent where
\begin{eqnarray}
	b_{l\lambda} &=& \int_0^1 h_{\lambda}\,\mu P_l\,d \mu \\
	\alpha_{\mathrm{T}\lambda} &=& {\partial \ln F_{\lambda}^+ \over
	\partial\ln\mathrm{T}_{\rm eff}} \;; \qquad
	\alpha_{g \lambda} = {\partial \ln F_{\lambda}^+ \over
	\partial\ln g_{e}} \\
	\beta_{\mathrm{T}\lambda} &=& {\partial\ln b_{l\lambda} \over
	\partial\ln\mathrm{T}_{\rm eff}} \;; \qquad
	\beta_{g \lambda} = {\partial\ln b_{l\lambda} \over
	\partial\ln g_{e}} \\
	f_{\mathrm{T}} &=& \Bigg|{\delta\mathrm{T}_{\rm eff}\over 
	\mathrm{T}_{\rm eff}}\Bigg|
	\;;\qquad
	f_{g} = \Bigg|{\delta g_{e}\over g_{e}}\Bigg|
\end{eqnarray}

\noindent $f_{\mathrm{T}}$ and $f_{g}$ are the relative amplitudes of local 
effective temperature and gravity variations for a normalized radial 
displacement at the photosphere, $\phi^T$ is the phase difference between the 
relative effective temperature variation and the relative radial displacement
and $\sigma$ is the pulsation frequency. $a$ corresponds to the real amplitude 
of the relative radial displacement, $P_l^m$ is the associated Legendre 
function and $i$ the inclination angle between the stellar rotation axis and 
the observer line of sight.

One appropriate way to test multicolor theoretical predictions is to construct 
phase-amplitude diagrams corresponding to well-chosen combinations of 
photometric bands. In these diagrams the theoretical results corresponding to 
modes of different spherical degrees $\ell$ occupy well differentiated regions.
 This enables the identification of $\ell$ by searching for the best fit 
between theory and observations. As is shown in Sect. \ref{deltascut}, the 
non-adiabatic quantities $f_{\mathrm{T}}$, $f_{g}$ and $\phi^T$, which play a 
major role in Eq. (\ref{deltaml}), are highly sensitive to
 the characteristics of the convective envelope. The confrontation between the 
theoretical and observed amplitude ratios and phase differences can thus 
be used to constrain the physical conditions of this convection zone.

\section{Applications to $\delta$ Scuti stars}
\label{deltascut}

\begin{figure}
\centering
	\resizebox{\hsize}{!}{\includegraphics{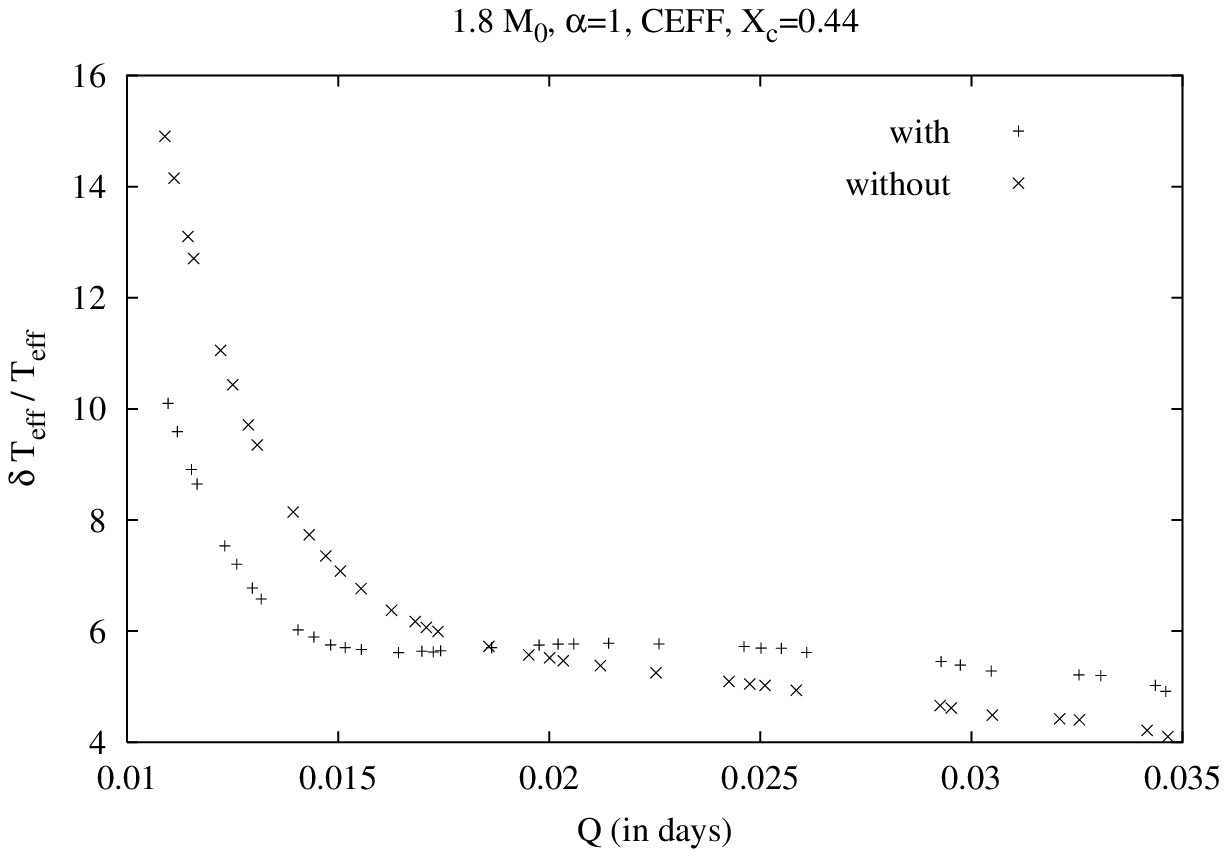}}
     	\resizebox{\hsize}{!}{\includegraphics{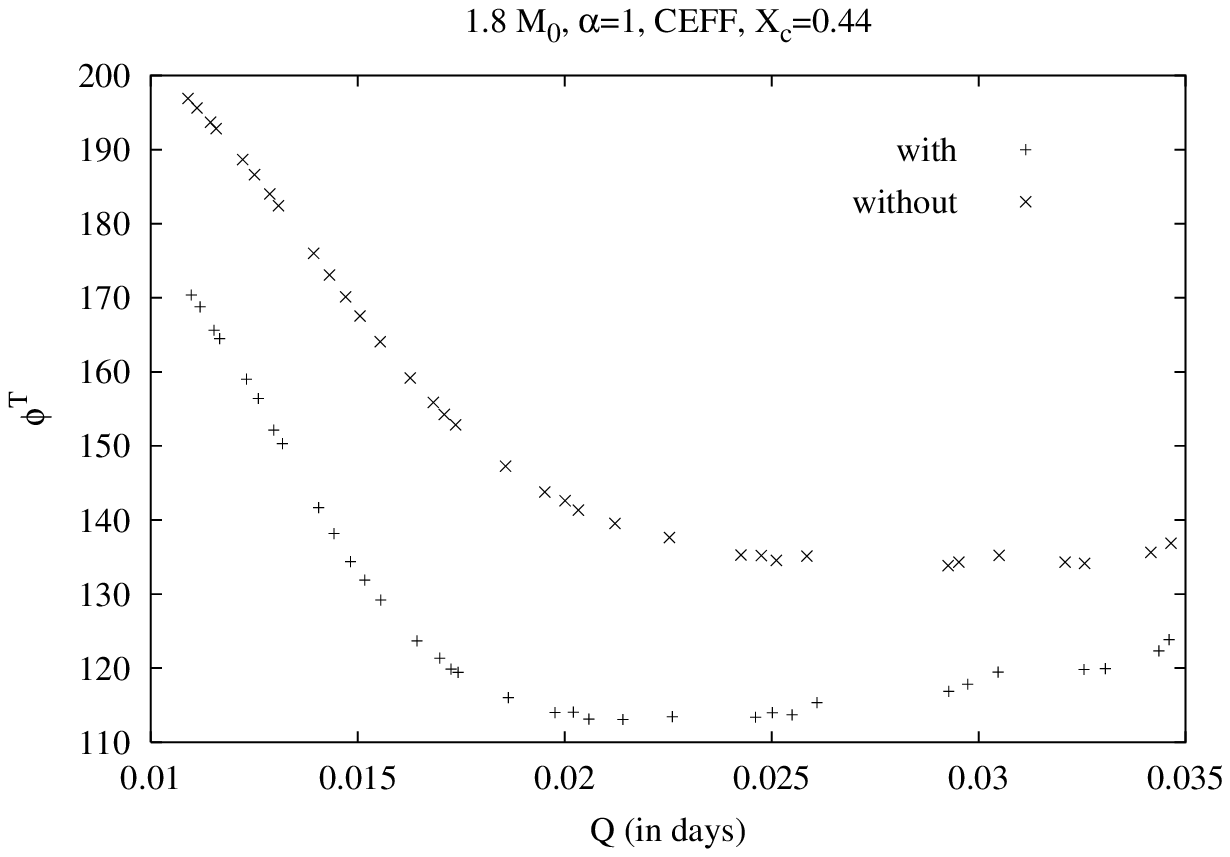}}
	\caption{Non-adiabatic quantities 
          $|\delta\mathrm{T}_{\rm eff} / \mathrm{T}_{\rm eff}|$ (top)
          and $\phi^T$ (bottom), as a function of the pulsation constant $Q$ 
          (in days) for different modes with spherical degrees $l=0,1,2,3$.  
          A model of a $1.8M_{\sun}$ is solved with $X_c=0.44$,
          a MLT parameter $\alpha =1$ and the CEFF equation of state. 
          Results obtained ``with'' ($+$) and 
          ``without'' atmosphere ($\times$) in the non-adiabatic 
          treatment are compared.}
\label{con_sin}
\end{figure}

As a prerequisite to the presentation of the ``with'' atmosphere treatment 
results, of particular interest is that the differences be analized between the
 non-adiabatic observables appearing as a consequence of ``with'' and 
``without'' atmosphere descriptions. In Fig. \ref{con_sin} 
these differences are displayed for a $1.8 M_{\sun}$ model in the middle of the
 evolutionary phase ($X_c=0.44$) towards the exhaustion hydrogen in the core.
For the relative amplitude of the local effective temperature 
variation, the effect of introducing the atmospheric treatment is slight, 
though significant, especially around the overstable modes (between Q=0.02
and Q=0.012 days in this model). However, the phase lag $\phi^T$ is 
more sensitive to this treatment for all frequencies and displays a mean 
difference of $30\degr$ for the overstable modes. As of the latter modes, 
these quantities increase with the frequency. It should be noted that the 
differences in $|{\delta g_e\over g_e}|$ between the ``with'' and ``without'' 
atmosphere models are negligible. Furthermore, it should be stressed that the 
frequencies obtained using both treatments are almost identical, i.e., the only
difference between the two treatments appears in the outermost layers, and this
has no effect on the frequency results. In Fig. \ref{con_sin} it can be seen 
that the non-adiabatic results for a fixed model are independent of the 
spherical degree $l$ for p-modes.

\begin{figure}
\centering
	\resizebox{\hsize}{!}{\includegraphics{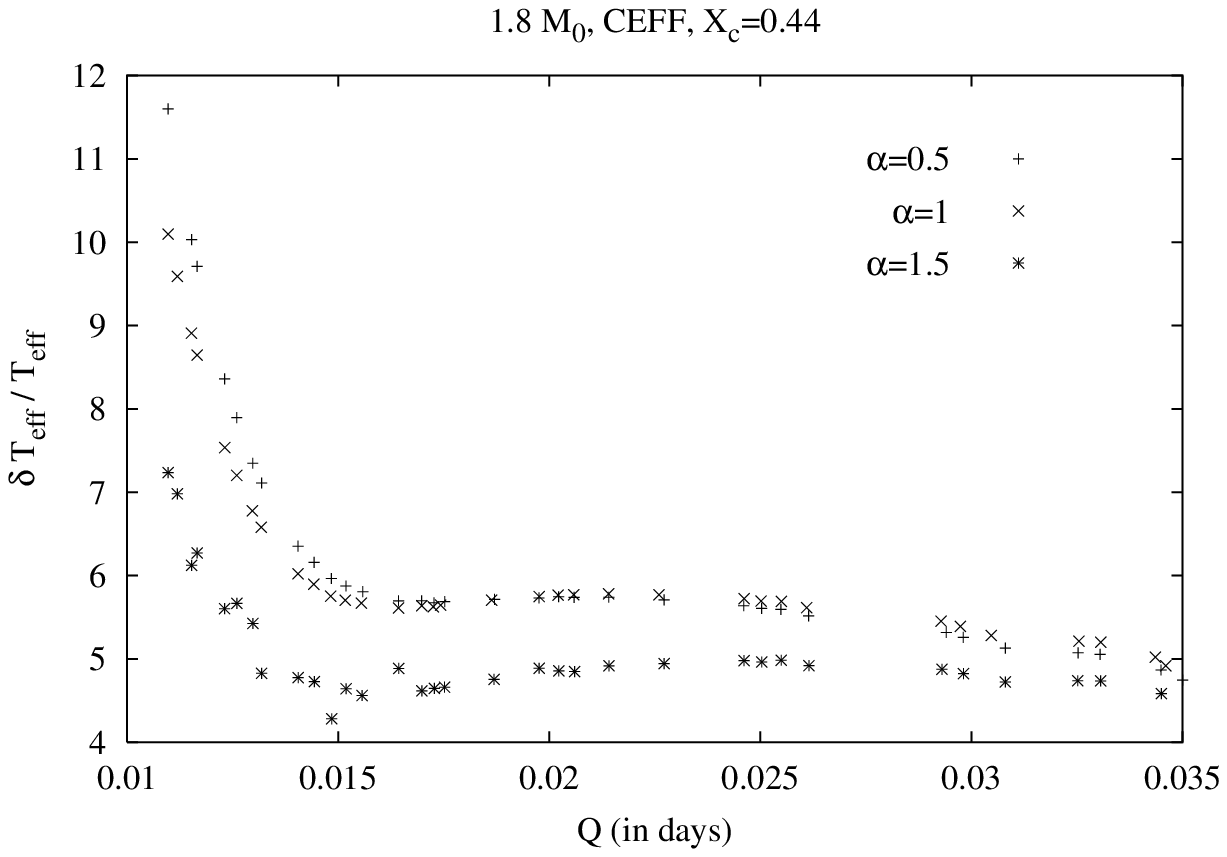}}
	\resizebox{\hsize}{!}{\includegraphics{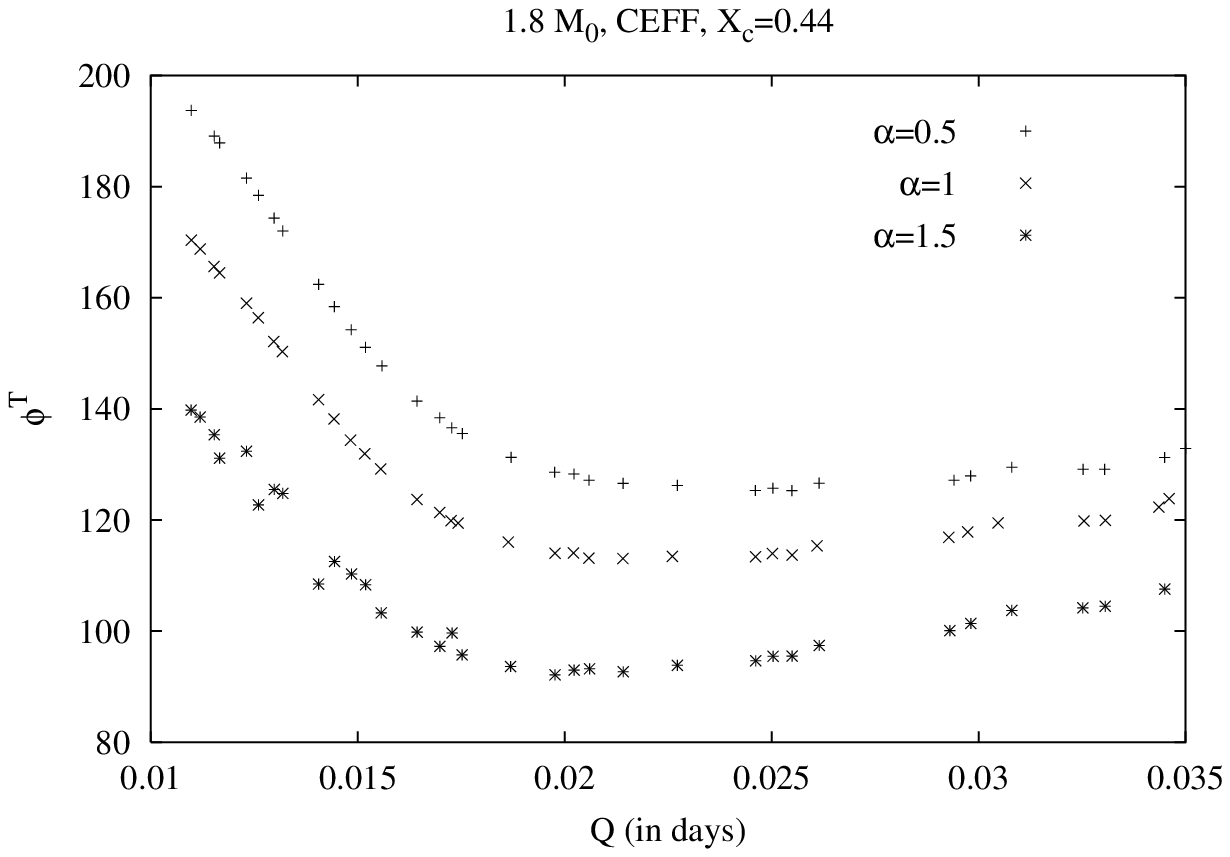}}
	\caption{Non-adiabatic quantities 
          $|\delta\mathrm{T}_{\rm eff} / \mathrm{T}_{\rm eff}|$ (top)
          and $\phi^T$ (bottom), as a function of the pulsation constant $Q$ 
          (in days) for different modes with spherical degrees $l=0,1,2,3$.  
          Models of $1.8M_{\sun}$ are solved with $X_c=0.44$
          and with different values for the MLT parameter: $\alpha =0.5,1,1.5$.
          Note the sensitivity of the non-adiabatic results with respect 
          to $\alpha$.}
\label{obs_alfa}
\end{figure}

\begin{figure}
\centering
	\resizebox{\hsize}{!}{\includegraphics{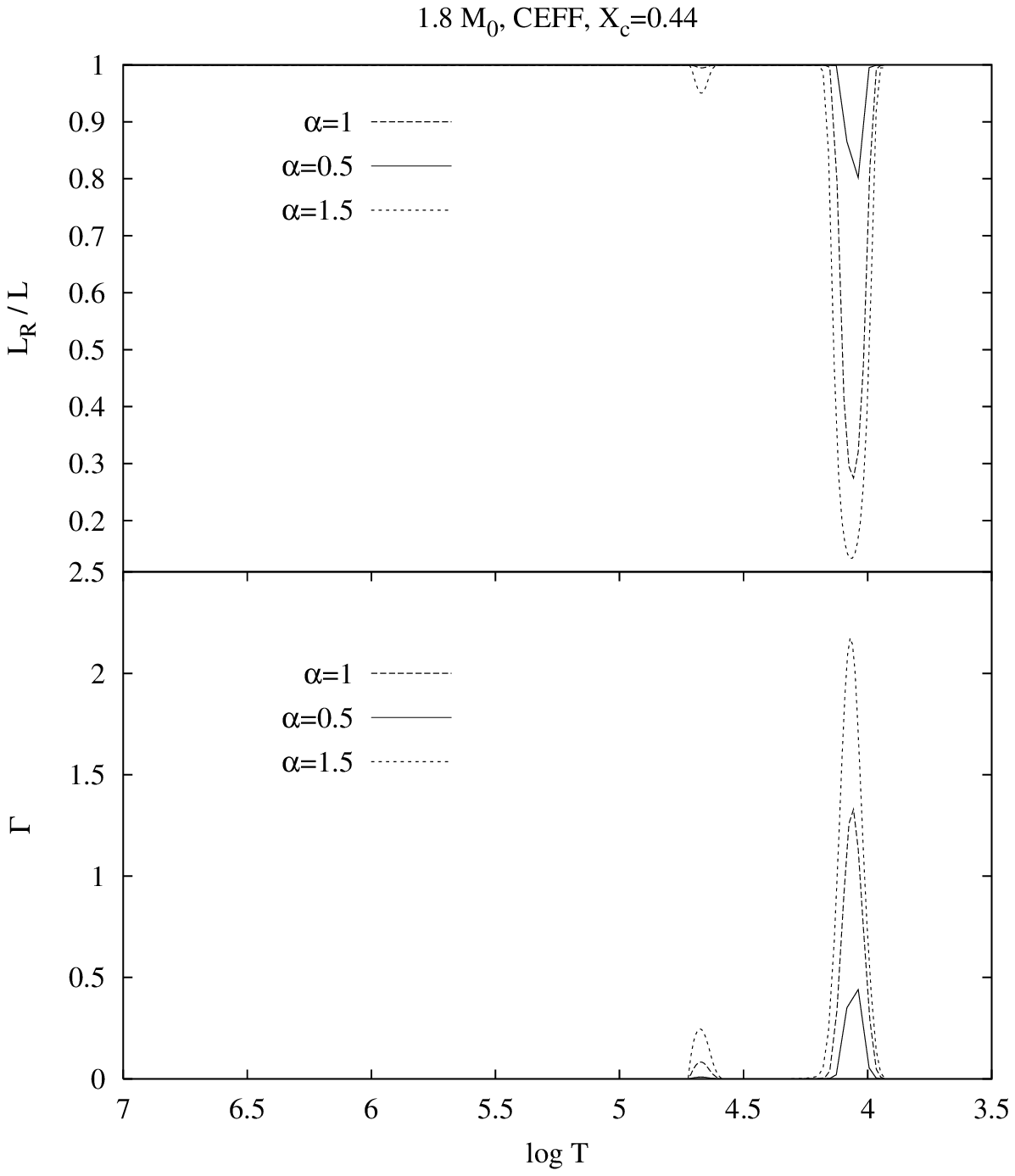}}
	\resizebox{\hsize}{!}{\includegraphics{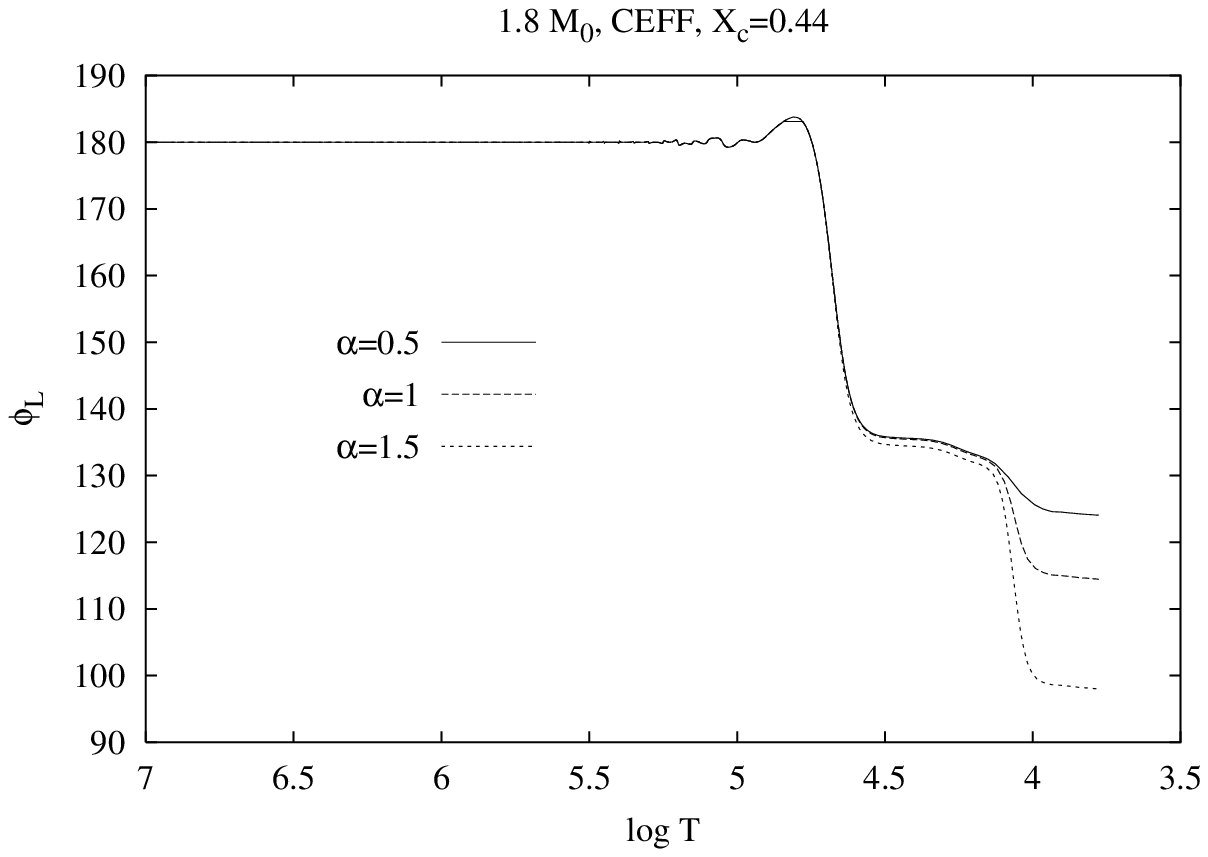}}
	\caption{Radiative luminosity over total luminosity (top panel),
convective efficiency (middle panel) and the luminosity phase lag
$\phi_L=\phi\big({\delta L\over L}\big)-\phi\big({\xi_r\over r}\big)$
(bottom panel) as a function of the logarithm of temperature,
for the fundamental radial mode. $1.8M_{\sun}$ models are solved with 
$X_c=0.44$, and three different values of the MLT parameter: 
$\alpha =0.5,1,1.5$. Note the differences appearing in the superficial
convection zone as $\alpha$ values rise.}
\label{efic_phase}
\end{figure}

Only pulsation models ``with'' atmosphere will be considered in what follows.
In Fig. \ref{obs_alfa} non-adiabatic results are compared as obtained for 
models of $1.8M_{\sun}$ and $X_c=0.44$, and with different values for the MLT 
parameter ($\alpha =0.5,1,1.5$).  The non-adiabatic 
quantities $|\delta\mathrm{T}_{\rm eff} / \mathrm{T}_{\rm eff}|$ and  $\phi^T$ 
are significantly affected by the value of $\alpha$, especially the 
phase lag. A more detailed description of the sensitivity of the non-adiabatic 
results to $\alpha$ is given in Fig. \ref{efic_phase}. This figure displays 
the ratio between radiative and total luminosity (top), the convective 
efficiency (middle)  and the luminosity phase lag 
$\phi_L=\phi\big({\delta L\over L}\big)-\phi\big({\xi_r\over r}\big)$ (bottom 
panel), all three as a function of the logarithm of temperature. Note that 
$\phi_L$  closely follows $\phi^T$ at the photosphere throughout Eq. 
(\ref{deltatt}).

\begin{figure}
\centering
     	\resizebox{\hsize}{!}{\includegraphics{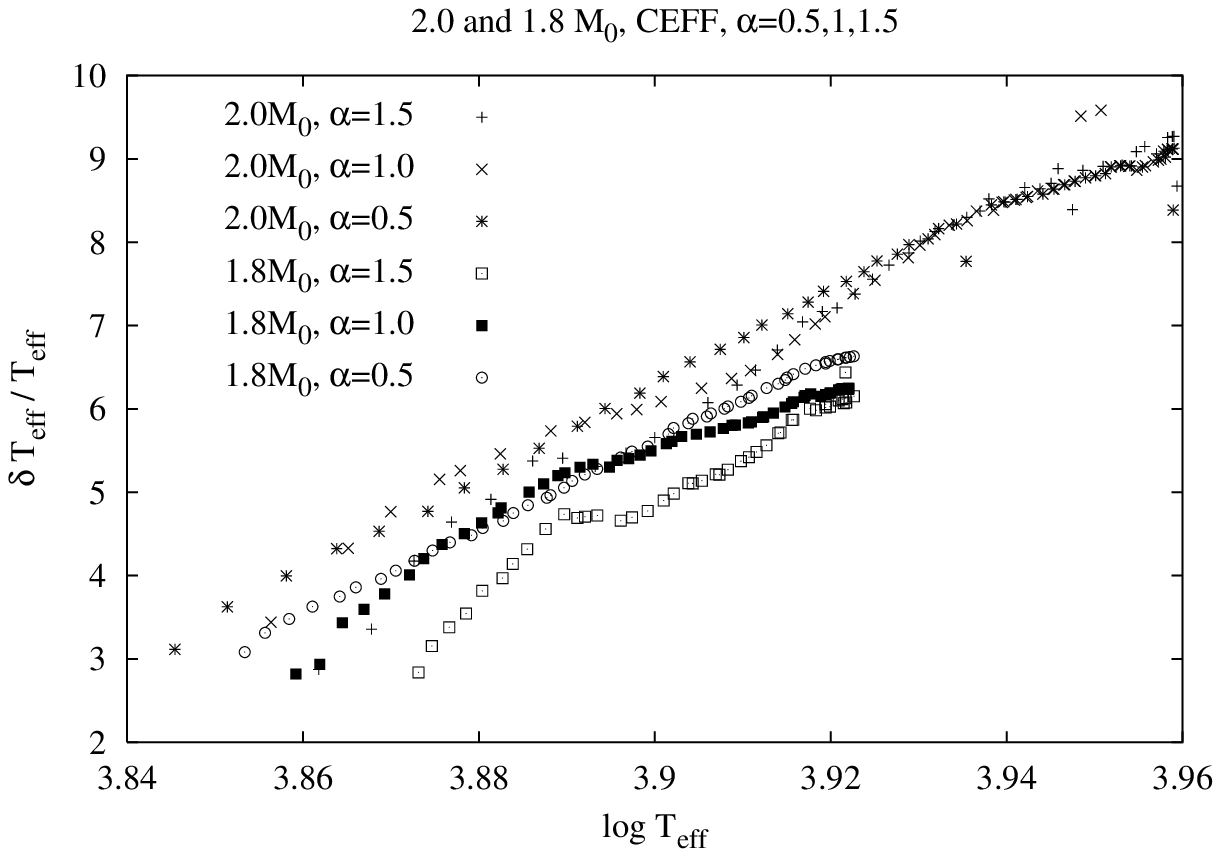}}
	\resizebox{\hsize}{!}{\includegraphics{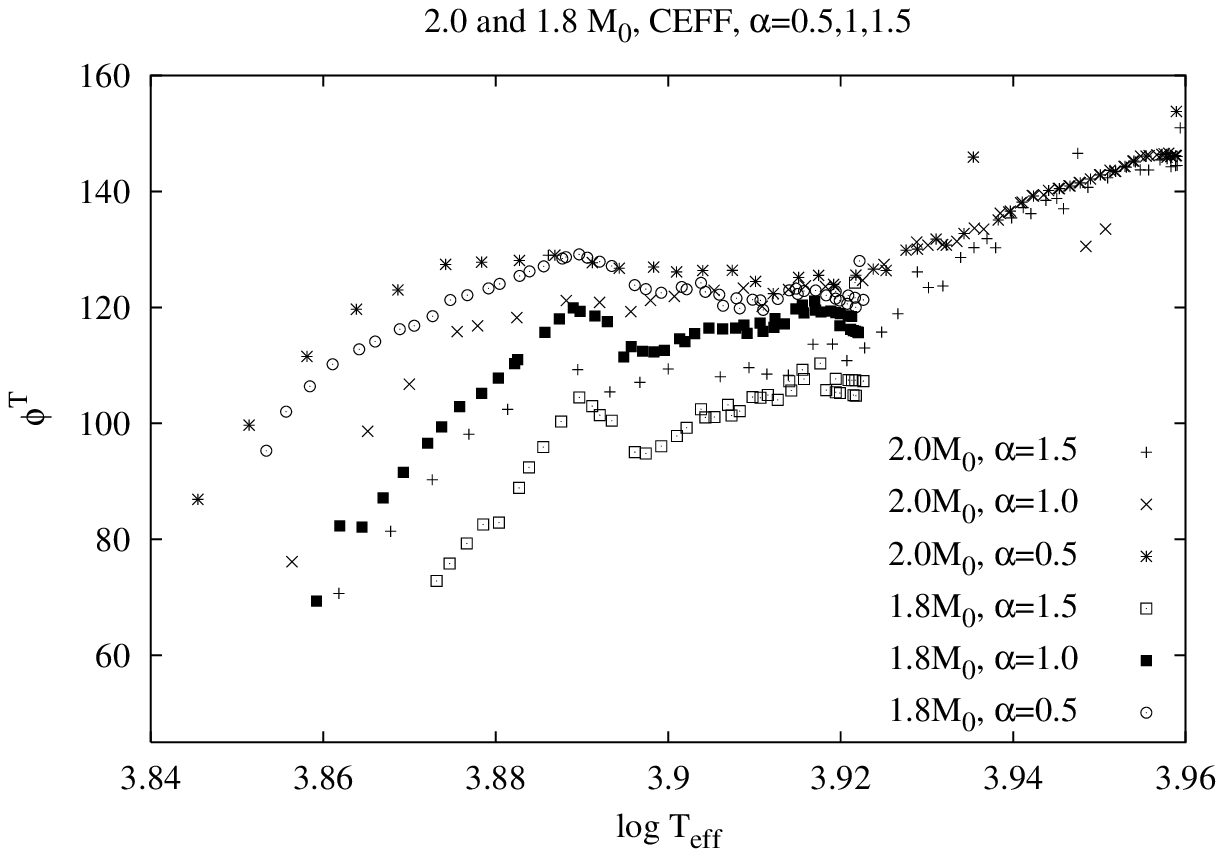}}
	\caption{Non-adiabatic quantities 
$|\delta\mathrm{T}_{\rm eff} / \mathrm{T}_{\rm eff}|$ (top) and $\phi^T$ 
(bottom) as a function of $\log T_{eff}$ for the fundamental radial mode in 
two complete tracks of 2.0$M_{\odot}$ and 1.8$M_{\odot}$ stars for three 
different values of the MLT parameter $\alpha=$1.5,1 and 0.5}
\label{phase_teff_teff}
\end{figure}

The bottom panel of Fig. \ref{efic_phase} shows that along the 
radius of the star there are two zones in which a phase lag is introduced: the 
first located in the partial ionization zone of HeII and the second, 
in the surface convection zone (partial HI and HeI ionization).
In this panel it can also be seen that the the phase lag sensitivity to 
$\alpha$ appears in the surface convection zone. 

The convective efficiency in MLT, defined as

\begin{equation}
	\Gamma=\Bigg[{4\over 9}\bigg({c_p\kappa p\rho c_s\alpha^2\over 
	9\sigma_{steff} T^3g\sqrt{2\Gamma_1}}\bigg)^2
	(\nabla_{rad}-\nabla)\Bigg]^{1\over 3}
\label{efic}
\end{equation}

\noindent (Cox \cite{COX}), and the convection zone temperature gradient are 
both directly linked to $\alpha$. Phase-lags originate 
in the energy conservation equation due to the introduction of
explicitly imaginary parts. For a radial mode, and freezing the convective 
luminosity, the equations are: 

\begin{equation}
	i\sigma T\delta S=\delta \epsilon_N-{d\delta L_R\over dM_r}
\label{energ_conserv}
\end{equation}

where $M_r$ is the mass enclosed in a sphere of radius $r$, and

\begin{equation}
\label{dlr}
	{\delta L_R\over L_R}  = 
		-{\delta\kappa\over\kappa}+4{\xi_r\over r}+4{\delta T\over T}+
		{d{\delta T\over T}/ d\ln r\over d\ln T/ d\ln r}
\end{equation}

Phase-lags originate mainly through the interplay between the different terms
of Eq. (\ref{dlr}), which affect the right hand side of Eq. 
(\ref{energ_conserv}) only when the thermal relaxation time is sufficiently 
small. A first phase-lag is introduced in the partial ionization zone of HeII 
(bottom panel of Fig. \ref{efic_phase}). The partial ionization produces an 
opacity bump and a considerable decrease in the adiabatic exponents. This 
significantly affects $\delta\kappa / \kappa$. A second source of 
phase lag occurs in the surface convective zone (partial ionization zone of HI 
and HeI). In this case variations in $\alpha$ primarily influence the 
temperature gradient and the size of the convective zone. The latter two then 
affect the phase-lag throughout the above equations.

As is shown in Fig. \ref{est_tau}, the non-adiabatic results are highly 
sensitive to the evolutionary phase of the star. As the star evolves in the HR 
diagram, changes in these quantities are larger for the relative variations of 
the effective temperature than for the phase lags. Changes in the evolution 
phase do not qualitatively affect the non-adiabatic results in the same manner 
as do changes in $\alpha$. In the latter case, only the characteristics of the 
convective zone are influenced. In contrast, changes in the evolution phase 
have an effect on both the characteristics of the superficial convective zone 
and the position of the HeII partial ionization zone.

\begin{figure}
\centering
	\resizebox{\hsize}{!}{\includegraphics{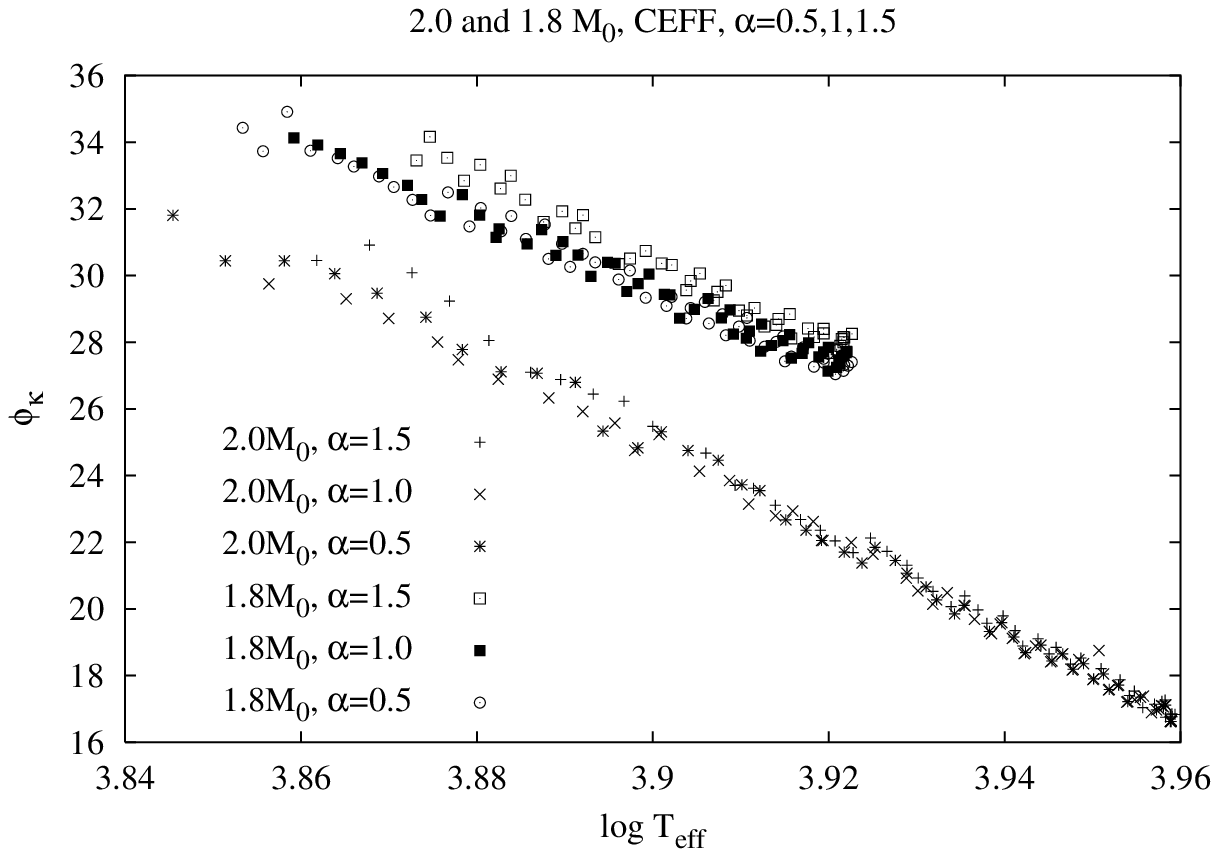}}
     	\resizebox{\hsize}{!}{\includegraphics{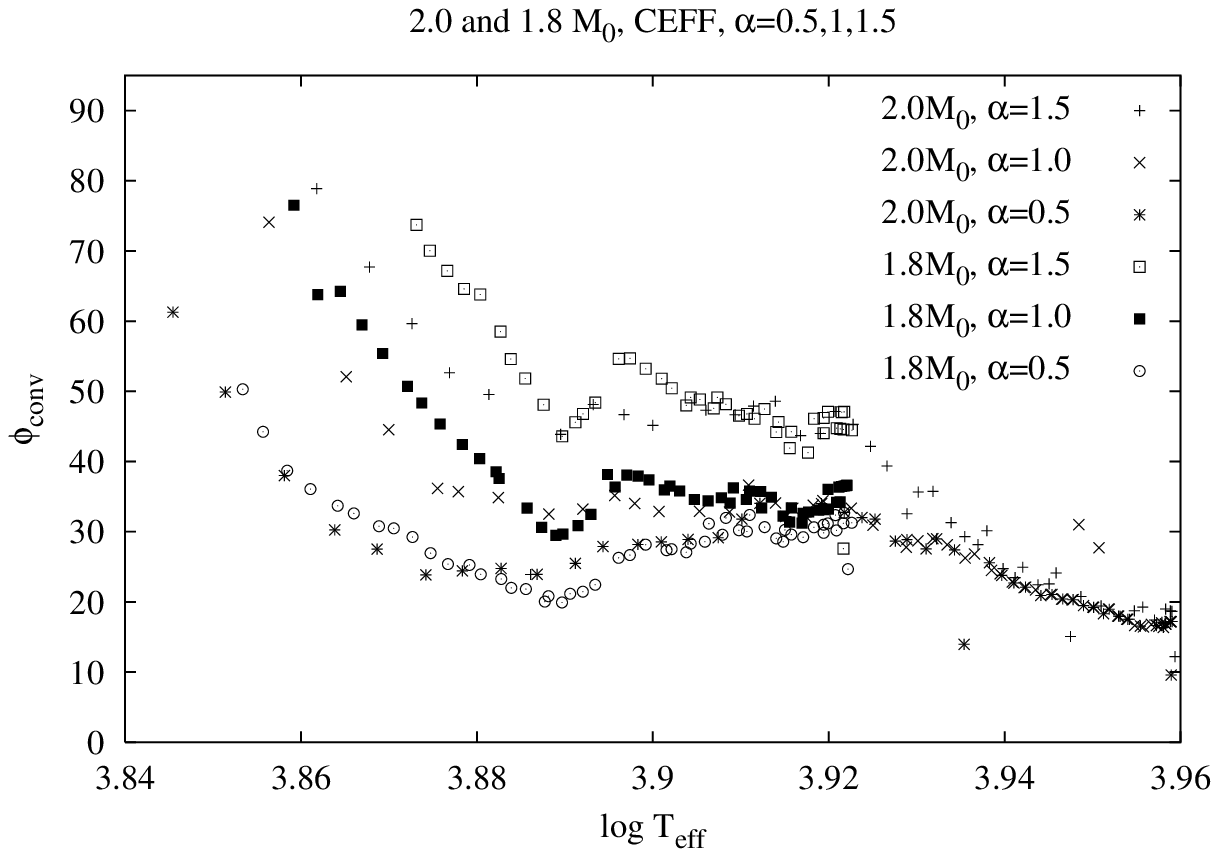}}
	\caption{$\kappa$-phase $\phi_{\kappa}$ (top) and convective-phase 
$\phi_{conv}$ (bottom) as a function of $\log T_{eff}$ for the fundamental 
radial mode in two complete tracks of 2.0$M_{\odot}$ and 1.8$M_{\odot}$ stars 
for three different values of the MLT parameter $\alpha=$1.5,1.0 and 0.5}
\label{phase_kap_conv_teff}
\end{figure}

\begin{figure}
\centering
	\resizebox{\hsize}{!}{\includegraphics{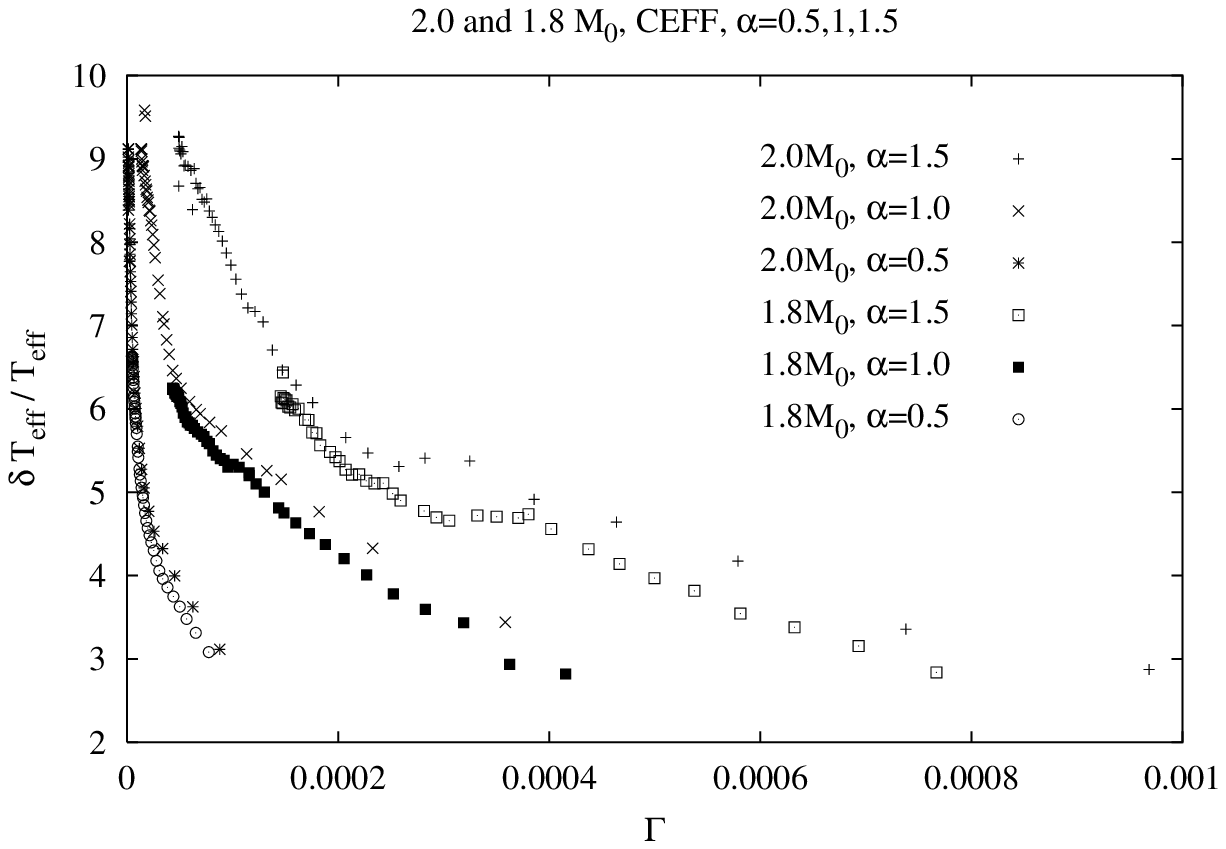}}
	\caption{$|\delta\mathrm{T}_{\rm eff} / \mathrm{T}_{\rm eff}|$
as a function of the integral of the convective efficiency $\Gamma$ for the 
fundamental radial mode in two complete tracks of 2.0$M_{\odot}$ and 
1.8$M_{\odot}$ stars for three different values of the MLT parameter 
$\alpha=$1.5,1.0 and 0.5}
\label{teff_efic}
\end{figure}

Fig. \ref{phase_teff_teff} is obtained by plotting the phase lag and 
$|\delta\mathrm{T}_{\rm eff} / \mathrm{T}_{\rm eff}|$ versus 
$\log \mathrm{T}_{\rm eff}$ for the fundamental radial mode of two complete 
tracks of 2.0$M_{\odot}$ and 1.8$M_{\odot}$ stars. Three values of 
the MLT parameter $\alpha$ (1.5,1.0 and 0.5) have been used. When stellar 
temperatures are high enough to originate a negligible external 
convective zone, the values are independent of $\alpha$; this only occurs for 
the 2.0$M_{\odot}$ models presented. For cooler models, however, convection
 becomes more efficient, thus producing different phase lags and relative 
effective temperature variations for similar effective temperatures (Balona \& 
Evers \cite{balev}).

Fig. \ref{efic_phase} refers to the behaviour of these phase lags. Each step 
in the figure makes it possible to view the phase lag, $\phi^T$, as 
the sum of two phases ($\phi_L=\phi_{\kappa}+\phi_{conv}$). The latter can be
referred to as the $\kappa$-phase ($\phi_{\kappa}\equiv$ phase difference 
between $180\degr$ and $\phi_L$ (at $\log T=4.5$)) and the
convective-phase ($\phi_{conv}\equiv$ phase difference between $\phi_L$ (at 
$\log T=4.5$) and $\phi_L$ (at $\log T=\log T_{eff}$)). Fig. 
\ref{phase_kap_conv_teff} is obtained by plotting both versus $\log T_{eff}$.

The $\kappa$-phase lag $\phi_{\kappa}$ displays a behaviour which is directly 
related to the position of the model in the HR diagram and which is independent
of the $\alpha$ parameter. Therefore, $\phi_{\kappa}$ shows the contribution of
only one of the phase sources, the opacity variation, since convection is 
not efficient in this part of the star.

The contribution to the convection zone phase, in this case $\phi_{conv}$, is 
in turn a sum of two phase sources. For hot models the $\kappa$ driving 
mechanism is not yet efficient enough to introduce a phase lag, and the 
behaviour of $\phi_{conv}$ depends mainly on the convection treatment. When the
 convective layers are not efficient, models with different values of $\alpha$ 
produce the same values of $\phi_{conv}$. At $\log T_{eff}\approx 3.91$ the 
convective layers become sufficiently efficient to distinguish between models 
using different values of $\alpha$, independently of the mass. From 
$\log T_{eff}=3.88$ to cooler models, an increase is observed in the value of 
this phase lag, again displaying different behaviours for different masses and 
values of $\alpha$. This is accounted for by the fact that, within this range 
of temperature, the $\kappa$ driving mechanism in the HI and HeI ionization 
zone becomes efficient enough to make a significant contribution to 
$\phi_{conv}$.

Fig. \ref{teff_efic} is obtained by displaying 
$|\delta\mathrm{T}_{\rm eff} / \mathrm{T}_{\rm eff}|$ versus the 
integral of the convective efficiency, as defined in eq. (\ref{efic}) for all 
models. The behaviour of these quantities can be distinguished 
independently of the mass of the model, which in this case is exclusively a 
function of the evolution phase and $\alpha$.

\section{Applications for multicolor photometry}

\begin{figure}
\centering
\rotatebox{270}{
	\resizebox{!}{\hsize}{\includegraphics{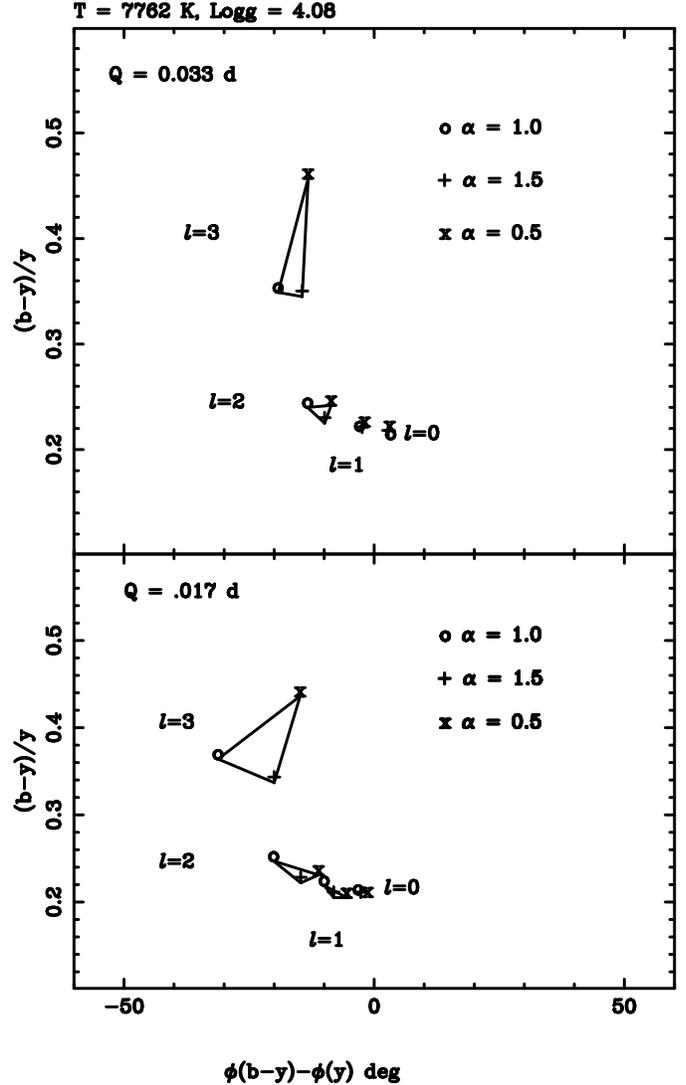}}}
	\caption{The top panel shows theoretical predictions for two specific 
Str\"omgren photometric bands ( ($b-y$) and $y$) for a given theoretical model 
using three MLT $\alpha$ parameters in the fundamental radial mode regime 
(pulsation constant near 0.033 days). The 3rd overtone regime 
(pulsation constant near 0.017 days) is shown in the bottom panel.}
\label{multi_1}
\end{figure}

All of the above-mentioned calculations have a direct influence on the phase 
difference - amplitude ratio diagrams. Now phase lags, as well as relative 
variations in $|\delta\mathrm{T}_{\rm eff} / \mathrm{T}_{\rm eff}|$ and  
$\delta g_e/g_e$ can be used to overcome the uncertainties in previous 
phase-ratio color diagrams. In Garrido (2000) these discrimination diagrams 
were made using parametrized values for departures from adiabaticity and phase 
lags. The only remaining degree of freedom is now the choice of the MLT 
$\alpha$ parameter in order to describe the convection. Therefore,
discrimination diagrams depend only on this parameter, as is 
shown in Fig. \ref{multi_1} where theoretical predictions are plotted for two
specific Str\"omgren photometric bands (($b-y$) and $y$) with a given 
theoretical model using three different MLT $\alpha$ parameters in the 
fundamental radial mode regime (pulsation constant near 0.033 days) and in the 
3rd overtone regime (near 0.017 days). 

A clear separation between the $l$-values exists for periods around the 
fundamental radial. Similar behaviour is found for other modes in the proximity
 of the 3rd radial overtone, although for these shorter periods some 
overlappings start to appear at the lowest $l$-values. They also show the same 
trend as in the fundamental radial mode: high amplitude ratios for low MLT 
$\alpha$ and the spherical harmonic $l=3$. A more detailed description of these
 diagrams using different models and photometric bands will be given in a 
forthcoming paper.

\section{Conclusions}

This paper uses the CESAM package to generate equilibrium models. A new 
non-adiabatic pulsation code has been developed and applied to the study of 
$\delta$ Scuti stars. This code takes into account the stellar atmosphere
 in two ways: 1) as the boundary layer of the star
(``without'' atmosphere); in this case photospheric observables 
such as amplitudes and phases of both $T_{eff}$ and gravity, reflect the 
boundary conditions we impose; and 2) particular consideration is given to  
pulsation treatment (``with'' atmosphere), thus extending the star beyond the 
photosphere. As a consequence of applying the ``with'' atmosphere treatment, 
photospheric observables become determinable as a solution to a set of 
differential equations. These equations have been described by Dupret et al. 
(\cite{MA02}).

The non-adiabatic code presented in this paper enables the determination of the
 photometric observables $|\delta T_{eff}/T_{eff}|$, $|\delta g_e/g_e|$ and 
$\phi^T$. The results of the ``without'' atmosphere approach strongly depend on
 the choice of external boundary conditions. In contrast, the ``with'' 
atmosphere approach only implies those uncertainties brought about by the 
physical assumptions imposed in order to obtain the differential equations.

The comparison of the results generated by both treatments shows that the 
values for $|\delta T_{eff}/T_{eff}|$ are relatively similar in most of the 
spectrum. However, differences become notable for shorter periods, in which 
``without'' atmosphere solutions are larger than those of the ``with'' 
atmosphere treatment. Differences in $\phi^T$ are significant and have a mean 
value of approximately $30\degr$ within the range of the periods 
studied. The non-adiabatic results presented here are highly sensitive to the 
characteristics of the superficial convective zone, parametrized by using
the mixing length parameter $\alpha$. In particular, we have shown that
there are two regions in which the phase lag originates. A first phase lag 
takes place in the partial ionization zone of HeII, where the $\kappa$ 
mechanism drives the oscillations. This phase lag is very
sensitive to the evolution phase. A second one occurs in the 
convective envelope (partial ionization zone of HI and HeI) and is
sensitive mainly to $\alpha$. Though to a lesser extent, this phase lag is 
also sensitive to the evolutionary phase because of the $\kappa$ driving 
mechanism and the size of the convection zone change.

In this paper it is shown that theoretical photometric amplitude ratios and 
phase differences are very sensitive to non-adiabatic calculations. 
The treatment and location of the outer boundary conditions, here
referred to as ``with'' and ``without'' atmosphere, have been shown
to give significantly different theoretical predictions. Consequently, our 
improved non-adiabatic treatment of the atmosphere will
probably enable more accurate photometric mode identifications. Our
results confirm the recent theoretical results of Daszy\'nska-Daszkiewicz 
et al. (\cite{Dasz03}) which show
that phases and amplitudes in different colours are affected by the 
choice of the $\alpha$ parameter, although the authors calculated the model 
atmospheres in the Eddington approximation. 
By determining the best fit between theory and observations, it may thus be 
possible to constrain the MLT parameter $\alpha$.

\begin{acknowledgements}
This work was partialy financed by program ESP 2001-4528-PE. The authors are
also grateful to Marco Bettini for thorough editing of the original English
manuscript.
\end{acknowledgements}



\appendix

\section{Stellar interior equations and inner boundary conditions}

These equations have been developed following Unno et al. 
(\cite{Unno}). Eigenfunctions 
${1\over gr}\big({p^\prime\over\rho}+\Phi^\prime\big)$ and 
${\delta S\over c_p}$ have been replaced by ${\delta P_g\over P_g}$ and 
${\delta T\over T}$
to adapt the equations to the atmospheric ones. This was done to make the
transition in the connecting layer as smooth as possible. Having introduced 
these modifications, the nonadiabatic {\bf nondimensional} equations for the 
stellar interior are

\begin{eqnarray}
	x{dy_1 \over dx} & = & 
		\Big({l\,(l+1) \over c_1 \omega^2}-3 \Big)\, y_1 +
\nonumber
\\
	&&
		\Big({l\,(l+1) \over c_1 \omega^2}-V_g-\nabla_{\mathrm{ad}}\,V
		\,\upsilon_{\mathrm{T}}\Big)\, {\beta \over V}\, y_2 +
		{l\,(l+1) \over c_1 \omega^2}\, y_3 +
\nonumber
\\
	&&
		\Bigg[{4 \over 3}{\mathrm{aT^4} \over \mathrm{PV}}
		\Big({l\,(l+1) \over c_1 \omega^2}-V_g-\nabla_{\mathrm{ad}}\,V
		\,\upsilon_{\mathrm{T}}\Big) + \upsilon_{\mathrm{T}}\Bigg]\, 
		y_5
\label{eq1}
\\
	x{dy_2 \over dx} & = &
		{V \over \beta}\, \Bigg( c_1 \omega^2 - \mathrm{U}+
		\Big(4-{l\,(l+1) \over c_1 \omega^2}\Big)\Big(1-
		{4 \over 3}{\mathrm{aT^4} \over \mathrm{P}}\nabla\Big)\Bigg)
		\, y_1 +
\nonumber
\\
	&&
		\Bigg[{V \over \beta}\,\Big(1-
		{4 \over 3}{\mathrm{aT^4} \over \mathrm{P}}\nabla\Big)-
		{l\,(l+1) \over c_1 \omega^2}
\nonumber
\\
	&&
		-{4 \over 3}{\mathrm{aT^4} \over 
		\mathrm{P}}\Big[V\, \nabla \big( \nabla_{\mathrm{ad}} 
		\kappa_{\mathrm{S}}-\kappa_{\mathrm{ad}} \big)
		-\nabla{l\,(l+1) \over c_1 \omega^2}\Big]\Bigg]\, y_2 +
\nonumber
\\&&
		{V \over \beta}\,{l\,(l+1) \over c_1 \omega^2}\Big(
		{4 \over 3}{\mathrm{aT^4} \over \mathrm{P}}\nabla-1\Big)\,y_3 -
		{V \over \beta}\, y_4 -
\nonumber
\\
	&&
		{4 \over 3}{\mathrm{aT^4} \over \mathrm{P\beta}}\Bigg[
		4\nabla+{l\,(l+1) \over c_1 \omega^2}+V\,\nabla\big(4-
		\kappa_{\mathrm{S}}\big)
\nonumber
\\
	&&
		+{4 \over 3}{\mathrm{aT^4} \over 
		\mathrm{P}}\Big[V\, \nabla \big( \nabla_{\mathrm{ad}} 
		\kappa_{\mathrm{S}}-\kappa_{\mathrm{ad}} \big)
		-{l\,(l+1) \over c_1 \omega^2}\Big]\Bigg]\, y_5 +
\nonumber
\\&&
		{V \over \beta}\, {4 \over 3}{\mathrm{aT^4} \over \mathrm{P}}
		\nabla\,y_6
\label{eq2}
\\
	x{dy_3 \over dx} & = &
		\big(1-\mathrm{U}\big)\,y_3+y_4
\label{eq3}
\\
	x{dy_4 \over dx} & = &
		\mathrm{U}\big(\mathrm{A}^*+V_g\big)\, y_1 +
		\mathrm{U}\,\beta\Big({1 \over \Gamma_1}+\upsilon_{\mathrm{T}}
		\nabla_{\mathrm{ad}}\Big)\, y_2 +
		l\,(l+1)\,y_3 - 
\nonumber
\\
	&&
		\mathrm{U}\,y_4 +
		\mathrm{U}\Bigg[{4 \over 3}{\mathrm{aT^4} \over \mathrm{P}}
		\Big({1 \over \Gamma_1}+\upsilon_{\mathrm{T}}
		\nabla_{\mathrm{ad}}\Big)-\upsilon_{\mathrm{T}}\Bigg]\, y_5
\label{eq4}
\\
	x{dy_5 \over dx} & = &
		V\,\nabla\Big(4-{l\,(l+1) \over c_1 \omega^2}\Big)\, y_1 +
		\beta\Big(V\,\nabla\big(\nabla_{\mathrm{ad}}\kappa_{\mathrm{S}}
		-\kappa_{\mathrm{ad}}\big)-
\nonumber
\\
	&&
		\nabla{l\,(l+1) \over c_1 \omega^2}
		\Big)\,y_2 -
		V\,\nabla{l\,(l+1) \over c_1 \omega^2}\,y_3 +
		\Bigg[V\,\nabla\big(4-\kappa_{\mathrm{S}}\big)+
\nonumber
\\
	&&
		{4 \over 3}
		{\mathrm{aT^4} \over \mathrm{P}}\Big(V\,\nabla\big(\nabla_
		{\mathrm{ad}}\kappa_{\mathrm{S}}-\kappa_{\mathrm{ad}}\big)
		-\nabla{l\,(l+1) \over c_1 \omega^2}\Big)\Bigg]\, y_5 -
\nonumber
\\
	&&
		V\,\nabla\,y_6
\label{eq5}
\\
	x{dy_6 \over dx} & = &
		l\,(l+1)\Big({c_3 \over c_1 \omega^2}-1\Big)\, y_1 +
\nonumber
\\
	&&
		\beta\Bigg[c_3\big(\varepsilon_{\mathrm{ad}}-\varepsilon_
		{\mathrm{S}}\nabla_{\mathrm{ad}}\big)+{l\,(l+1) c_3 \over 
		c_1 \omega^2 V}+\mathrm{i}\omega c_4 \nabla_{\mathrm{ad}}
		\Bigg]\,y_2 -
\nonumber
\\
	&&
		l(l+1){c_3 \over c_1 \omega^2}\,y_3 +
		\Bigg[{4 \over 3}{\mathrm{aT^4} \over \mathrm{P}}\Big(
		c_3\big(\varepsilon_{\mathrm{ad}}-\varepsilon_{\mathrm{S}}
		\nabla{\mathrm{ad}}\big)+
\nonumber
\\
	&&
		{l\,(l+1) c_3 \over 
		c_1 \omega^2 V}+\mathrm{i}\omega c_4 \nabla_{\mathrm{ad}}
		\Big)+c_3\varepsilon_{\mathrm{S}}-
\nonumber
\\
	&&
		{l(l+1) \over \nabla\,V}-
		\mathrm{i}\omega c_4\Bigg]\, y_5 -
		{d\,\ln\,L_R \over d\,\ln\,r}\, y_6
\label{eq6}
\end{eqnarray}

\noindent where $x=r/R$, and $R$ is the photometric radius. The rest of the 
variables follow the definitions given in Unno et al. (\cite{Unno}):

\begin{eqnarray}
	V_g = {V \over \Gamma_1} = -{1 \over \Gamma_1}{d\ln \mathrm{P}
	\over d\ln r} = {gr \over c^2} 
	\;\;\;\;\;\;\;
	\mathrm{U} = {d\ln M_r \over d\ln r} = {4\pi\rho r^3 \over M_r}
\nonumber
\\
	c = \Big({\Gamma_1\mathrm{P} \over \rho}\Big)^{1/2}
	\;\;\;\;\;\;\; 
	c_1 = {x^3 \over M_r/M} 
	\;\;\;\;\;\;\; 
	\omega^2 = {\sigma^2R^3 \over GM} 
\end{eqnarray}
\begin{eqnarray}
	\beta = {P_g \over \mathrm{P}} \;\;\;\; \nabla_{\mathrm{ad}} = 
	\Big({\partial\ln T \over \partial\ln\mathrm{P}}\Big)_{\mathrm{S}}
	\;\;\;\; \nabla = {d\ln T \over d\ln\mathrm{P}}
	\;\;\;\;A^* = {rN^2 \over g}
\vspace{0.1cm}
\nonumber
\\
	\upsilon_{\mathrm{T}} = -\Big({\partial\ln\rho \over 
	\partial\ln\mathrm{T}}\Big)_{\mathrm{P}}
	\;\;\;\; c_3 = {4\pi r^3\rho\varepsilon_{\mathrm{N}} \over L_R}
\vspace{0.1cm}
\nonumber
\\
	c_4 = {4\pi r^3\rho\mathrm{T}c_p \over L_R}\sqrt{{GM \over R^3}}
	\;\;\;\;
	\kappa\,(\,\mathrm{or}\;\varepsilon_{\mathrm{N}})_{\mathrm{ad}} = 
	\Big({\partial\ln\kappa\,(\,\mathrm{or}\;\varepsilon_{\mathrm{N}}) 
	\over \partial\ln\mathrm{P}}\Big)_{\mathrm{S}}
\vspace{0.1cm}
\nonumber
\\
	\kappa\,(\,\mathrm{or}\;\varepsilon_{\mathrm{N}})_{\mathrm{S}} = c_p
	\Big({\partial\ln\kappa\,(\,\mathrm{or}\;\varepsilon_{\mathrm{N}}) 
	\over \partial\mathrm{S}}\Big)_{\mathrm{P}}
\end{eqnarray}

\noindent $N$ is the Brunt-V\"ais\"al\"a frequency and, $M$ the total mass of 
the star, $\kappa$, is the opacity and, $\varepsilon_{\mathrm{N}}$,  is the 
energy generation rate.\\

The interior boundary conditions are common to both pulsational treatments,
being that the stellar interior is solved using the same equations. In our 
formalism, these conditions are:
\begin{enumerate}
	\item The mechanical boundary condition
\vspace{0.1cm}
	\begin{equation}
		\big(c_1\omega^2-l\big)\,y_1-{l\,\beta\over V}\,y_2-l\,y_3
		-l{4\over 3}{a\mathrm{T}^4 \over \mathrm{P}V}\,y_5 = 0
	\label{intmbc}
	\end{equation}
\vspace{0.1cm}
	\item The potential boundary condition
\vspace{0.1cm}
	\begin{equation}
		l\,y_3-y_4 = 0
	\label{intpbc}
	\end{equation}
\vspace{0.1cm}
	\item The thermodynamical boundary condition
\vspace{0.1cm}
	\begin{eqnarray}
		-\nabla_{\mathrm{ad}}\,\beta\, y_2+\Big(1-{4\over 3}
		{a\mathrm{T}^4 \over \mathrm{P}}\nabla_{\mathrm{ad}}\Big)
		\,y_5 = 0\\
	\label{inttbc}
\nonumber
	\end{eqnarray}
\end{enumerate}


\begin{thebibliography}{}

\bibitem[2001]{Are02} Arentoft, T., Sterken, C., Handler, G. et al., 2001,
A\&A, 374, 1056

\bibitem[1999]{balev} Balona, L.A., Evers, E.A., 1999, MNRAS 302, 349

\bibitem[1979a]{Bal79a} Balona, L.A., Stobie, R.S., 1979a, MNRAS 189, 627

\bibitem[1979b]{Bal79b} Balona, L.A., Stobie, R.S., 1979b, MNRAS 189, 649

\bibitem[1999]{Bre99} Breger, M., Pamyatnykh, A. A., Pikall, H. et al., 1999,
A\&A 341, 151

\bibitem[2002]{Bre02a} Breger, M., Garrido, R. and Handler, G., 2002, MNRAS,
329, 531

\bibitem[1999]{Bre02b} Breger, M., Handler, G., Garrido, R. et al., 1999,
A\&A, 349, 225

\bibitem[1980]{COX} Cox, J.P., 1980,
  Theory of Stellar Pulsation,
  Princeton Univ. Press, Princeton

\bibitem[1994]{cug} Cugier, H., Dziembowski, W., Pamyatnykh A., 1994, 
  A\&A 291, 143

\bibitem[2002]{Dasz02} Daszy\'nska-Daszkiewicz, J., Dziembowski, W.A., 
Pamyatnykh, A.A.,  Goupil, M.-J., 2002,  A\&A 392, 151 

\bibitem[2003]{Dasz03} Daszy\'nska-Daszkiewicz, J., Dziembowski, W.A., 
Pamyatnykh, A.A., 2003, A\&A {\bf 407, 999}

\bibitem[2002]{MA02} Dupret, M.-A., De Ridder, J., Neuforge, C., Aerts, C., 
  and Scuflaire R., 2002, A\&A 385, 563

\bibitem[2003]{MA03} Dupret, M.-A., De Ridder, J., De Cat, P., Aerts, C., 
  Scuflaire, R., Noels, A., Thoul, A., 2003, A\&A 398, 677

\bibitem[1977a]{Dziem77a} Dziembowski, W., 1977a, 
  Acta Astron. 27, 95

\bibitem[1977b]{Dziem77b} Dziembowski, W., 1977b, 
  Acta Astron. 27, 203

\bibitem[2000]{Rafa00} Garrido, R., 2000,
  In: The 6th Vienna Workshop on $\delta\,$Scuti and related stars,
  eds. M. Montgommery, M. Breger, PASP Conference Series, 210, 67

\bibitem[1990]{Rafa90} Garrido, R., Garcia-Lobo, E., Rodriguez, E., 1990, 
  A\&A 234, 262

\bibitem[1982]{JCD} Christensen-Dalsgaard, J., 1982, MNRAS 199, 735

\bibitem[1993]{Kurucz} Kurucz, R.L., 1993,
  ATLAS9 Stellar Atmosphere programs and 2 km/s grids. Kurucz CDROM
  No 13 

\bibitem[1997]{Morel97} Morel, P., 1997, A\&A 124, 597

\bibitem[1990]{Pes90} Pesnell, W. D., 1990 ApJ 363, 227

\bibitem[1980]{Saio80} Saio, H. and Cox, J. P., 1980, ApJ 236, 549

\bibitem[2002]{Townsend02} Townsend, R., 2002, MNRAS 330, 855

\bibitem[1995]{FILOU} Tran Minh F. and Leon L., 1995, Phys. Pross. Ap., 219

\bibitem[1989]{Unno} Unno, W., Osaki, Y., Ando, H. et al., 1989, Nonradial
oscillation of stars, Univ. Tokyo Press, Tokyo

\bibitem[1988]{Wat88} Watson, R.D., 1988, Ap\&SS 140, 255

\end{thebibliography}
\end{document}